\documentclass[preprint,aip,floatfix,graphics]{revtex4-1}

\usepackage{amsmath}    
\usepackage{graphicx}   
\usepackage{verbatim}   
\usepackage{color}      
\usepackage{subfigure}  
\usepackage{hyperref}   
\usepackage{longtable}
\newcommand{\red}[1]{\textcolor{black}{#1}}

\begin{document}
\setlength{\LTcapwidth}{\linewidth}

\title{Coupled-cluster theory for atoms and molecules in strong magnetic fields}
\author{Stella Stopkowicz}\email{stella.stopkowicz@kjemi.uio.no}\affiliation{Centre for Theoretical and Computational Chemistry (CTCC), Department of Chemistry, University of Oslo, P.O.Box 1033 Blindern, 0315 Oslo, Norway}
\author{J\"urgen Gauss}\affiliation{Institut f{\"u}r Physikalische Chemie, Universit\"at Mainz,
D-55099 Mainz, Germany}
\author{Kai K. Lange} \affiliation{Centre for Theoretical and Computational Chemistry (CTCC), Department of Chemistry, University of Oslo, P.O.Box 1033 Blindern, 0315 Oslo, Norway}
\author{Erik I. Tellgren} \affiliation{Centre for Theoretical and Computational Chemistry (CTCC), Department of Chemistry, University of Oslo, P.O.Box 1033 Blindern, 0315 Oslo, Norway}
\author{Trygve Helgaker}\affiliation{Centre for Theoretical and Computational Chemistry (CTCC), Department of Chemistry, University of Oslo, P.O.Box 1033 Blindern, 0315 Oslo, Norway}
\date{today}

\begin{abstract}
An implementation of coupled-cluster (CC) theory to treat atoms and molecules in finite magnetic fields is presented.
The main challenges for the implementation stem from the magnetic-field dependence in the Hamiltonian, or, more precisely, the appearance of the 
angular momentum operator, due to which the wave function becomes complex and which introduces a gauge-origin dependence.
For this reason, an implementation of a complex CC code is required together with the use of gauge-including atomic orbitals 
to ensure gauge-origin independence.  Results of coupled-cluster singles--doubles--perturbative-triples
(CCSD(T)) calculations are presented for atoms and molecules with a focus on the dependence of correlation  and binding energies 
on the magnetic field.
\end{abstract}

\maketitle
\section{Introduction}
The investigation of atoms and molecules in strong magnetic fields has become a topic of high interest in the 
recent years.\cite{SchmelcherRev1,Duncan,Lai,KaiScience,Toru}
Even though strong fields of 1000\;T and more (note that one atomic unit, $B_0$, corresponds to $2.35 \times 10^5$\;T) 
cannot be generated and investigated on Earth, they occur in the atmospheres of magnetized white dwarf stars.\cite{review_white_dwarf}
Helium \cite{ExpevidenceHe,ExpevidenceHe2} and recently also hydrogen molecules\cite{ExpevidenceH2} have already been observed in such objects,
making the investigation of atoms and molecules in strong magnetic field not only interesting for understanding fundamental physical and chemical concepts but also important for astrophysics,\cite{review_white_dwarf} in particular, for the interpretation of observational spectra of white dwarfs and the determination of their magnetic field strength. 

While on Earth the strongest sustained laboratory fields are of the order of $10^{-4}B_0$, 
suggesting that the magnetic field can be treated as a perturbation, this is no longer true when increasing the field strength towards 1$B_0$.
In such a case, magnetic and Coulomb forces are of the same order of magnitude, making a non-perturbative treatment of the magnetic field essential. 
In a Science perspective article,\cite{zitatschmelcher} Schmelcher points out that
"\textit{the competition between the anisotropy-introducing magnetic field and the attractive and repulsive Coulomb forces is responsible for an enormous complexity and diversity of the microscopic behavior. Indeed, the existing investigations show that different excited states of a molecule, or ground states of similar molecules, can behave in a vastly different manner, and provide a first look at a largely unexplored area: the world of magnetized matter.}"
In fact, theoretical investigations have already led to new insights like the transition to diamagnetic behavior for paramagnetic closed-shell molecules \cite{eriklondon} and the discovery of a previously unknown perpendicular paramagnetic bonding mechanism.\cite{KaiScience}

As the study of atoms and, in particular, molecules in strong magnetic fields is still a rather uncharted territory and experiments are not possible, reliable theoretical investigations and accurate quantum-chemical calculations are required.  
So far, for the treatment of electron correlation, only full configuration-interaction (FCI) calculations have been reported---see, for example, 
Refs.~\onlinecite{HeliumFCI,LithiumFCI,KaiScience}, which due to their tremendous computational cost are restricted to very small systems 
with few electrons. 
It is therefore desirable to extend the capability of performing highly accurate quantum-chemical investigations to larger systems, as the increasing complexity of the systems studied may lead to new phenomena and new insights. 
Additionally, there is a great interest to formulate a density-functional theory (DFT) that properly takes into account the interaction with magnetic fields.\cite{CDFT1a,CDFT1b,BDFT,CDFT1c,Erik_DFT,DFThighmagnetic,Andy_CDFT}
For these approaches, it is mandatory to have benchmark values to assess the quality of the results and help to construct improved functionals.

In the present work, we report the formulation and implementation of coupled-cluster (CC) theory for the treatment of atoms and molecules in finite magnetic fields. 
In this way, we overcome the limitations of FCI theory, while retaining the ability to provide a highly-accurate treatment of molecules in strong magnetic fields. 
The focus of the present work is on CC theory with single, double and perturbative triple excitations (CCSD(T)),\cite{ccbook} 
considered the "gold standard" of quantum chemistry, shown in numerous cases to provide results of quantitative accuracy.

The paper is organized as follows. Having outlined the underlying theory in
Section \ref{theory}, we point out in Section \ref{implementation} the differences relative to regular, field-free CC implementations and comment on the validation of our implementation.
In Section\;\ref{applications}, we discuss applications, considering the helium, neon, fluorine, lithium, beryllium, and sodium atoms 
as well as the LiH and He$_3$ molecules.
The focus is here on the dependence of the total and the correlation energies on the magnetic field, on basis-set requirements \red{including a comparison with previous FCI results, as well as on} binding energies.

\section{Theory}
\label{theory}
The electronic Hamiltonian for a molecule with $N$ electrons in a uniform magnetic field takes the form
\\
\begin{align}
\label{Hinfield}
\hat H = \hat H_0 &+  \frac{1}{2} \sum_i^N \mathbf B \cdot \mathbf l^\mathrm{O}_i + \mathbf {B}\cdot \mathbf S \nonumber \\
& + \frac{1}{8} \sum_i^N \left( B^2 {r^\mathrm{O}_i}^2 - \left( \mathbf{B} \cdot \mathbf{r}_i^\mathrm{O}   \right)^2 \right)
&
\end{align}
with the usual field-free electronic Hamiltonian $\hat H_0$
composed of the kinetic-energy operator for the electrons and the potential-energy contributions of the electron--electron and 
electron--nucleus interactions.
In the additional terms, $\mathbf B$ is the magnetic field, $\mathbf S$ the total spin, $\mathbf r^\mathrm{O}_i = \mathbf r_i - \mathbf{O}$ the position vector for the $i$th electron with respect to the global gauge origin $\mathbf O$, and $\mathbf l^\mathrm{O}_i= -\mathrm i\mathbf{r}^\mathrm{O}_i \times \mathbf{\nabla}_i$ the canonical angular momentum.
The terms linear in $\mathbf B$ are called paramagnetic and constitute the orbital- and spin-Zeeman terms, respectively. 
The term quadratic in $\mathbf B$ is referred to as the diamagnetic term.  

In contrast to the field-free Hamiltonian usually used in quantum-chemical calculations, the Hamiltonian in \eqref{Hinfield} is gauge-origin dependent and leads to complex wave functions due to the presence of the electronic angular-momentum operator.
The spin-quantization axis is defined by the direction of the magnetic field. 

Turning to the electronic-structure problem in the presence of the magnetic field, we note that the
 derivation of CC energy and amplitude equations \cite{ccbook} follows exactly the same route as in the field-free case, with equivalent final equations.
We therefore give only a very brief overview over CC theory here, focusing on the key aspects.
As ansatz for the wave function, in CC theory an exponential form 
\begin{align}
\mid \Psi_\mathrm{CC}  \rangle  = \mathrm e^{\hat T} \mid \Phi_0 \rangle
\label{ansatz}
\end{align}
is chosen to ensure size extensivity.  The cluster operator 
\begin{align}
\hat T =& \hat T_1 + \hat T_2 + \dots +\hat T_N \\
=& \sum_{n=1}^N \left(\frac{1}{n!}\right)^2 \sum_{ij..ab} t_{ij..}^{ab..} \hat a_a^\dagger \hat a_i \hat a_b^\dagger \hat a_j \dots
\label{T2}
\end{align}
consists of the unknown amplitudes $t_{ij..}^{ab..}$ as well as products of particle creation and annihilation operators that
generate all possible excitations from the reference state $\Phi_0$, which in our case is the Hartree--Fock (HF) determinant $\Psi_\mathrm{HF}$.
In \eqref{T2}, the indices $i,j,\dots$ refer to occupied and $a,b,\dots$ to virtual spin orbitals.

To obtain equations to solve for the unknown amplitudes, the ansatz \eqref{ansatz} is inserted into the Schr\"odinger equation and the HF energy $E_\mathrm{HF}$ is subtracted. The equation is then premultiplied by $e^{-\hat T}$ and projected onto the excited determinants:
\begin{align}
\label{cceq}
\langle \Phi_I \mid \mathrm e^{-\hat T } \hat H_N e^{\hat T}\mid \Psi_\mathrm{HF} \rangle = 0
\end{align}
with  $\hat H_N = \hat H-E_\mathrm{HF} $.
Projection onto the HF reference determinant yields the CC correlation energy
\begin{align}
\langle \Psi_\mathrm{HF} \mid \mathrm e^{-\hat T } \hat H_N \mathrm e^{\hat T}\mid \Psi_\mathrm{HF} \rangle = E^\mathrm{CC}_{\mathrm{corr}} .
\end{align}
Approximate CC models are obtained by truncating the cluster operator at certain excitation levels. 
Choosing, for example, $\hat T =\hat T_1 + \hat T_2$ and projecting onto the singly and doubly excited determinants in \eqref{cceq}, 
we obtain the CCSD model.\cite{CCSD}
The additional perturbative treatment of triple excitations gives rise to the well-known CCSD(T) model.\cite{CCSD(T)}

Even though the final expressions look the same as in the field-free case, care has to be taken with respect to the permutational symmetries, as in the presence of a magnetic field all 
quantities---that is, integrals, intermediates, and amplitudes---become complex.

\section{Implementation}
\label{implementation}
A complex coupled-cluster code has been written in C++ as part of the LONDON program \cite{LONDON} following the intermediate 
formalism used by Stanton \emph{et al}.\cite{intermediates}$^,$\footnote{Note that the first integral in equation 2 in Ref. \onlinecite{intermediates} should be \unexpanded{$ \langle ab\mid\mid ij \rangle$}}
The non-linear equations for the amplitudes are solved using standard iterative schemes \cite{intermediates} as in the field-free case.
  
The program employs an unrestricted HF (UHF) reference to treat both closed- and open-shell species and
builds upon the existing functionality in LONDON for self-consistent-field (SCF) calculations and integral-evaluation over plane-wave/Gaussian type orbital (GTO) basis sets\cite{eriklondon}. 
LONDON uses gauge-including atomic orbitals (GIAOs)\cite{GIAO} to achieve gauge-origin invariance, which in a non-perturbative setting requires hybrid basis sets.
Recent work has also generalized the implementation, as well as the notion of gauge-origin invariance, to non-uniform magnetic fields.\cite{TELLGREN_JCP139_164118}.
Schemes for the computation of energies have been implemented at the CCSD\cite{CCSD} and CCSD(T) \cite{CCSD(T)} levels of theory.

The differences with respect to standard CC codes are that 
\begin{itemize}
   \item 
   all quantities, except energies and orbital energies, are complex;
   \item
there is less permutational symmetry for the two-electron integrals:
\begin{align}
\notag
\langle pq \mid rs \rangle  &= \langle rs \mid pq \rangle^* = 
\langle qp\mid sr \rangle = \langle sr\mid qp \rangle^* \\
&\ne (\langle ps\mid rq\rangle =\langle rq\mid ps \rangle^* = \langle qr\mid sp\rangle =\langle sp\mid qr\rangle^*
)
\end{align}
\item 
there is increased computational cost since, for
each quantity, the real and imaginary part needs to be stored and the operation count for multiplications is four times higher.
\end{itemize}
The code has been validated in the following way:
\begin{itemize}
\item 
For $\mathrm{B}=0$, amplitudes, intermediates, and final energies have been compared with the field-free CC code in the CFOUR program package.\cite{cfour}
\item
Magnetizabilities have been obtained using polynomial fitting to energies in the magnetic field and compared with magnetizabilities calculated 
analytically using CFOUR.\cite{Magnetiz} 
\item 
In cases where CCSD is equivalent to FCI, the CC results were compared with those obtained with the existing FCI code in LONDON.\cite{KaiScience,KaiFCI}
\item
For specific cases such as S states of atoms, the contribution from the magnetic field reduces to the diamagnetic term, 
which can be viewed as a confining harmonic potential in the directions orthogonal to the field and dealt with using standard CC codes. 
Similarly, such a treatment is possible for linear molecules in $\Sigma$ states with the field applied parallel to the molecular axis. 
\end{itemize}

\section{Applications}
\label{applications}
While there have been various accounts in the literature of what happens with atoms and molecules in (strong) magnetic 
fields (see, for example, Refs.\;\onlinecite{SchmelcherRev1,Duncan,Lai,KaiScience,Toru}), 
we focus here on the dependence of correlation and binding energies on the magnetic field.
We start by discussing the correlation energies in atoms and consider for that purpose helium, neon, fluorine, lithium, beryllium, and sodium. We then move on to molecules and discuss the correlation energy of LiH, thereby considering singlet and triplet states and different 
orientations with respect to the field.
Next, the basis-set requirements for correlation energies are discussed using as examples the Li $^2$P state, the $^1$D state of beryllium, the $^2$S, 
$^2$P, $^2$D, and $^2$F state of sodium as well as the LiH triplet state in perpendicular field.\footnote{Note that we use throughout this paper term symbols for the field-free case. Furthermore, for a given angular momentum and multiplicity we consider the state with lowest $M_S$ and $M_L$ value as it is most stable due to the spin-Zeeman and orbital-Zeeman contributions, respectively.}
\red{Additionally, we compare our results with FCI calculations from the literature\cite{HeliumFCI,HeliumFCIp,LithiumFCI,BerylliumCI,H2numbers,H2numbersPi,Schmelchercorr} for systems with few electrons. 
}
Finally, we investigate the effect of electron correlation on the binding energies in LiH and He$_3$, 
where the latter is particularly interesting as it only becomes bound in a strong magnetic field.\cite{ErikGradient}

All calculations have been carried out using Cartesian Gaussians employing GIAOs and uncontracted basis sets---namely, 
the uncontracted aug-cc-pCVQZ \cite{Dunningaug1,DunningPC} set unless stated otherwise.
The magnetic fields considered in this study range between $\text{B}=0$ and 1.6\,$B_0$, but we discuss mainly the results for fields up to 1\,$B_0$.
Geometries for LiH and He$_3$ were obtained numerically at the CCSD(T) level with uncontracted aug-cc-pVTZ \cite{Dunningaug1} basis sets via polynomial fitting to computed potential-energy curves. 
\subsection{Correlation energies for atoms}
\begin{figure*}
   \includegraphics{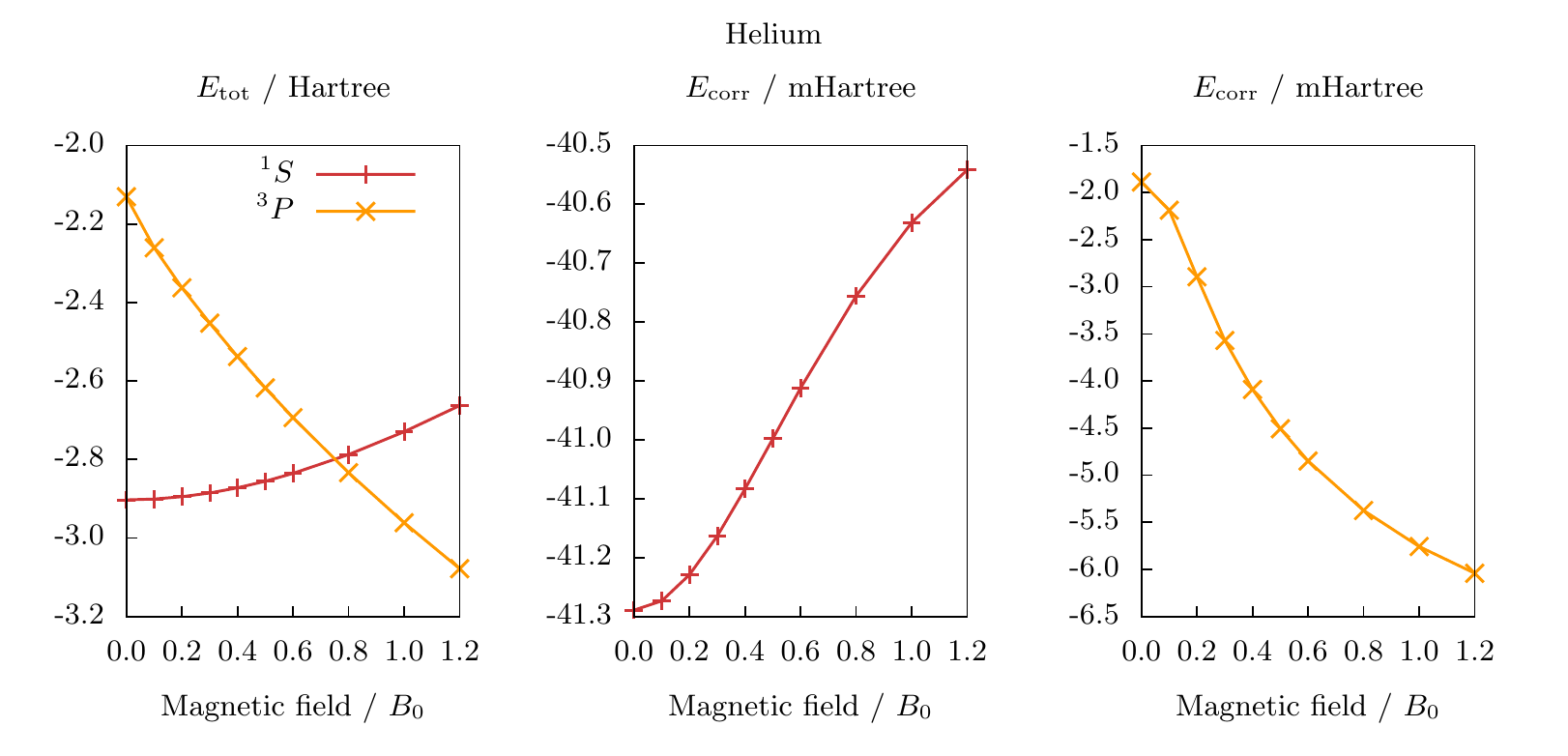}
   \caption{Total energies and correlation energies of helium as a function of the magnetic field. Left: Total energies of the helium $^1$S 
and the $^3$P state, middle: correlation energy of the $^1$S state, right: correlation energy of the $^3$P state. 
   Calculated at the CCSD(T) level with the uncontracted d-aug-cc-pVQZ basis set.
   }
   \label{He}
\end{figure*}

\noindent\textit{Helium}\\
$\;$ \\
Starting with the simplest system, the helium atom, we note that the energy of the ground state in the field-free case 
($^1$S state) increases diamagnetically when the magnetic field is turned on (see Figure \ref{He}). 
At a field of around 0.8\,$B_0$, the $^3$P triplet state becomes the ground state as its energy is paramagnetically lowered by both the orbital-Zeeman and the spin-Zeeman contributions (see also Refs.\;\onlinecite{HeliumFCI,HeliumFCI2}). 
The correlation energy for the singlet state is lowered in absolute magnitude with the field.
This can be explained in the following way:
electron correlation usually leads to a spatial expansion of the electronic distribution. 
There is a penalty for this expansion in the magnetic field due to a corresponding increase in the diamagnetic contribution. 
The latter is always positive and larger for spatially more extended systems---that is, for a field in $z$-direction:
\begin{align}
E^\mathrm{dia} =  \frac{1}{8} B_z^2 \langle \Psi \mid \sum_i(x_i^2+y_i^2) \mid \Psi \rangle
.
\end{align}
If the spatial extension in the correlated treatment is greater than at the HF level of theory, 
$E^{\mathrm{dia}}_{\mathrm{CC}} > E^\mathrm{dia}_\mathrm{HF}\;$
\footnote{
Note that this definition of spatial extension is not unique but serves here our purpose to understand the effect of the diamagnetic term in the Hamiltonian. 
In fact, measuring the spatial extension via other criteria can lead to a qualitatively different picture (see, for example, Ref. \onlinecite{Gillcorrelation}).
}, then the increase of the diamagnetic contribution reduces the correlation energy. 
Thus, electron correlation is hampered in the presence of a magnetic field. 

\begin{figure*}
   \includegraphics{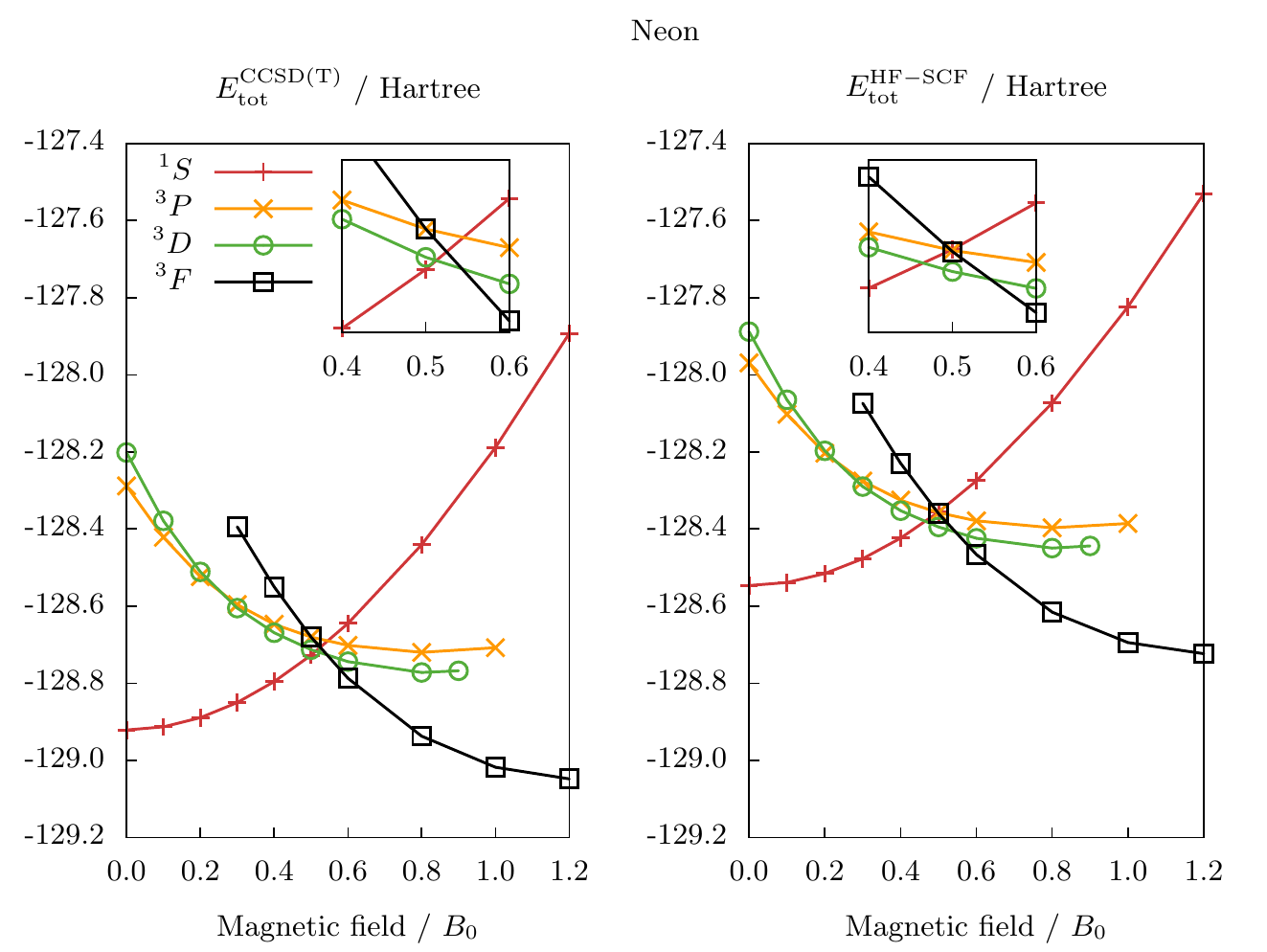}
   \caption{CCSD(T) and HF-SCF energies of the $^1$S, $^3$P, $^3$D, and $^3$F states of neon as a function of the magnetic field performed with the uncontracted aug-cc-pCVQZ basis set.
   }
   \label{Ne}
\end{figure*}

\begin{figure*}
   \includegraphics{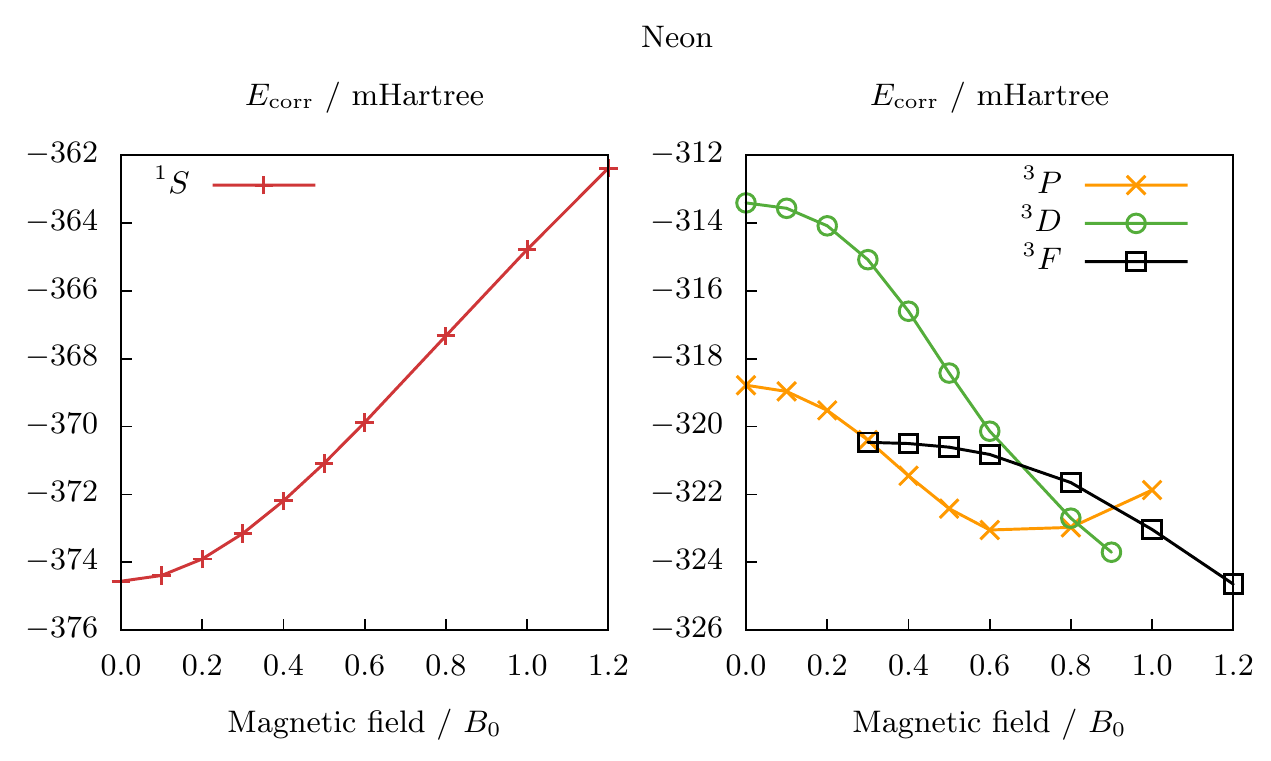}
   \caption{Correlation energy for the $^1$S, $^3$P, $^3$D, and $^3$F states of neon as a function of the magnetic field. 
   Calculated at the CCSD(T) level with the uncontracted aug-cc-pCVQZ basis set.}
   \label{Ne_corr}
\end{figure*}

\begin{figure*}
   \includegraphics{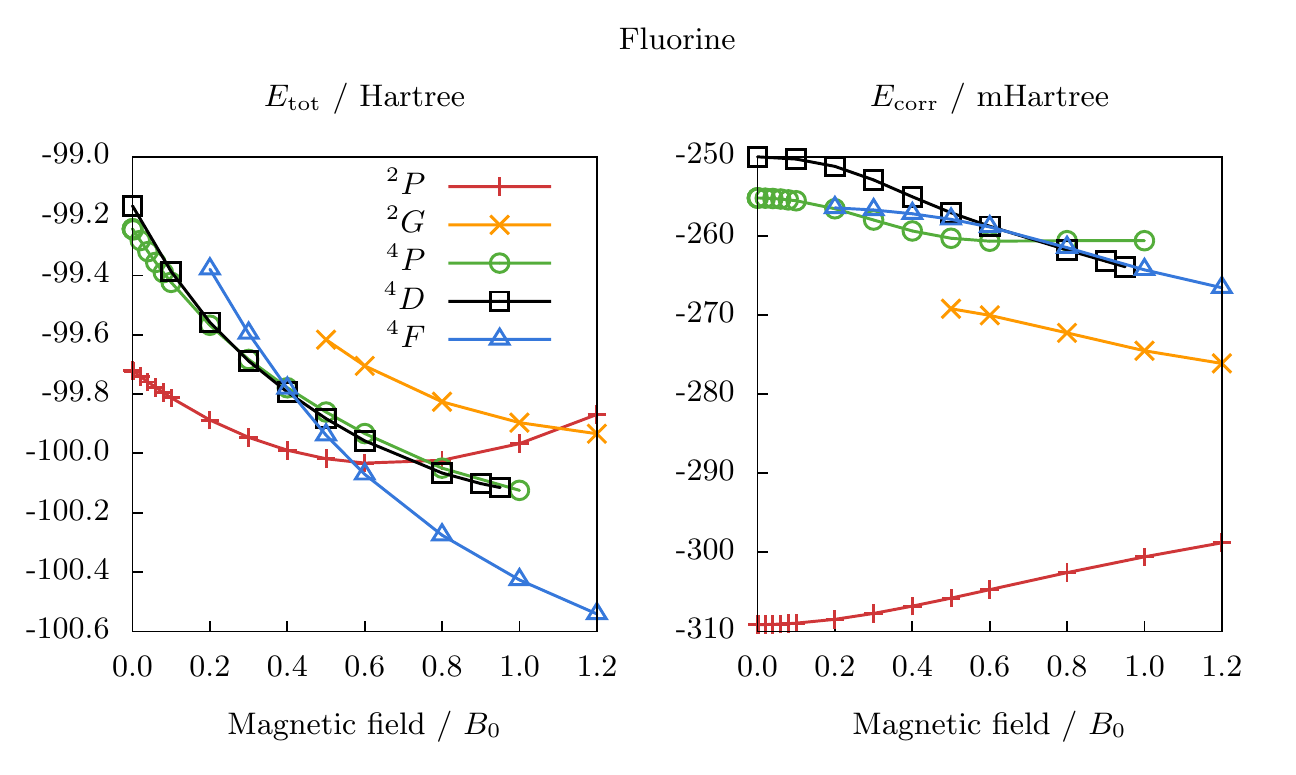}
   \caption{Total and correlation energies of the $^2$P, $^2$G, $^4$P, $^4$D, and $^4$F states of fluorine as a function of the magnetic field. 
   Calculated at the CCSD(T) level with the uncontracted aug-cc-pCVQZ basis set.}
   \label{F}
\end{figure*}

In the triplet state, a different picture emerges. 
First of all, electron correlation effects are here about one order of magnitude smaller since electrons of the same spin occupy different orbitals. 
Second, the correlation energy increases in absolute magnitude with the magnetic field.
In this case, an analysis of the field-free case shows that correlation leads to an expansion of the 1s orbital and a contraction of the 
higher-lying 2p orbital as an indirect effect, due to a reduced screening of the nuclear charge (see Table\;\ref{table_spatial}). 
In total, the electronic structure contracts, which is beneficial in the magnetic field. 
\linespread{0.5}
\begin{table}
\begin{footnotesize}
\caption{
Spatial extension (expectation values for $x^2$ and $y^2$, in bohr$^2$) for the 
occupied orbitals in case of a HF calculation and for the leading natural orbitals in 
case of a CCSD(T) calculation computed for He, Li, Be, and Ne. All
calculations have been performed with the uncontracted aug-cc-pVTZ basis set in the field-free limit. 
The occupation numbers of the natural orbitals are given in parentheses.
The entry `sum' which determines whether the total spatial extension is larger for HF or for CCSD(T) is calculated for HF as the sum of expectation values for one direction over all occupied orbitals, while for CCSD(T) we sum over all orbitals with the occupation numbers as weights.
The larger of the two contributions is marked in boldface.
}
\label{table_spatial}
\begin{tabular*}{\linewidth}{l@{\extracolsep\fill} c c c c}
\hline
\hline
Atom\;  &   State  &  Orbital &       HF    &           CCSD(T)    \\
\hline
He   &   $^1$S  &  1s    &      0.3950          &        {0.3967} (0.9919)\\
     &          &  sum   &      0.7900        &    \textbf{0.7976} \\        
\\
     &   $^3$P  &  1s    & 0.2794/0.2687 & 0.2803/0.2692 (0.9987)\\ 
     &          &  2p    & 3.6637/1.2212 & 3.6559/1.2186 (0.9987)\\
     &          & sum    & \textbf{3.9432/1.4900}      &   3.9356/1.4876     \\
\\
Li   &   $^2$S  &    1s($\alpha$) &   0.1480            &     0.1833  (0.9967) \\           
     &          &  1s($\beta$)    & 0.1496              &      0.1502  (0.9965) \\
     &          &  2s($\alpha$)   & 5.9092              &       5.8014  (0.9995) \\
     &          &  sum            &  \textbf{6.2068}      &       6.1344 \\ 
\\
     &   $^2$P  &  1s($\alpha$)   & 0.1499/0.1494     &  0.1512/0.1506 (0.9996) \\
     &          &  1s($\beta$)    & 0.1484/0.1487     &  0.1496/0.1499 (0.9973) \\
     &          &  2p($\alpha$)   &16.7116/5.5705     & 16.3845/5.4615 (0.9973) \\
     &          &  sum            &  \textbf{17.0098/5.8686}   & 16.7026/5.7816 \\
\\
Be  &   $^1$S &   1s     &      0.0777   &          0.0770   (0.9982) \\
    &          &   2s       &    2.8098     &        2.6493   (0.9106) \\
    &          &  sum       &  \textbf{5.7730}  &    5.4306  \\ 
\\
    &  $^3$P   &  1s($\alpha$)  & 0.0781/0.0778 &  0.0844/0.0832 (0.9982) \\ 
    &          &   1s($\beta$)  & 0.0779/0.0781 &  0.0781/0.0783 (0.9981) \\
    &          &  2s($\alpha$)  & 2.9443/2.3613 &  2.8698/2.3575 (0.9922)  \\
    &          &  2p($\alpha$)  & 6.1159/2.0279 &  6.1253/2.0429 (0.9923) \\
    &          & sum   &  \textbf{9.2162}/4.5452  &  9.1616/\textbf{4.5810}  \\   
\\
Ne  &   $^1$S   &  1s     &      0.0111                &    0.0113    (0.9997) \\
    &           &  2s     &      0.3222                &    0.3238    (0.9945) \\
    &           &  2p     &     0.7382/0.2462           &  0.7506/0.2514 (0.9890) \\
    &           & sum     &   3.1277                &   \textbf{3.2042}\\  
\hline
\hline
\end{tabular*}
\end{footnotesize}
\end{table}
\\ \\
\textit{Neon}
\\ \\
For neon, see Figure\;\ref{Ne}, the ground state is up to about B=0.5\,$B_0$ the $^1$S state, whose energy increases diamagnetically. 
Interestingly, HF calculations predict the $^3$D state to be the ground state at this field strength, while at the CCSD(T) level, the singlet state is still the lowest in energy.
For higher fields, the paramagnetically stabilized $^3$F state becomes the ground state. 
As for helium, the correlation energy for the singlet state is reduced in absolute terms, while for the triplet states the correlation energy 
increases (see Figure\;\ref{Ne_corr}). 
We note that, for the higher lying $^3$P state, the correlation energy also initially rises in absolute magnitude 
but starts to decrease around B=0.7\,$B_0$.
\\ \\
\textit{Fluorine}
\\ \\
Moving to open-shell systems, a similar picture emerges for fluorine as for the previously discussed systems (see Figure \ref{F}). 
The ground state is up to about 0.5 $B_0$ the $^2$P state, which is paramagnetically stabilized but exhibits a turning point 
at roughly 0.6 $B_0$, where the diamagnetic term starts to dominate. 
The correlation energy of this state is reduced in absolute magnitude with the field. 
The $^4$F state becomes the ground state beyond 0.6\,$B_0$, benefiting from a  much higher stabilization by both the spin- and the orbital-Zeeman term. 
The correlation energy rises with the magnetic field for this state, again due to the contraction of the electronic distribution at the CC level. 

\begin{figure*}
   \includegraphics{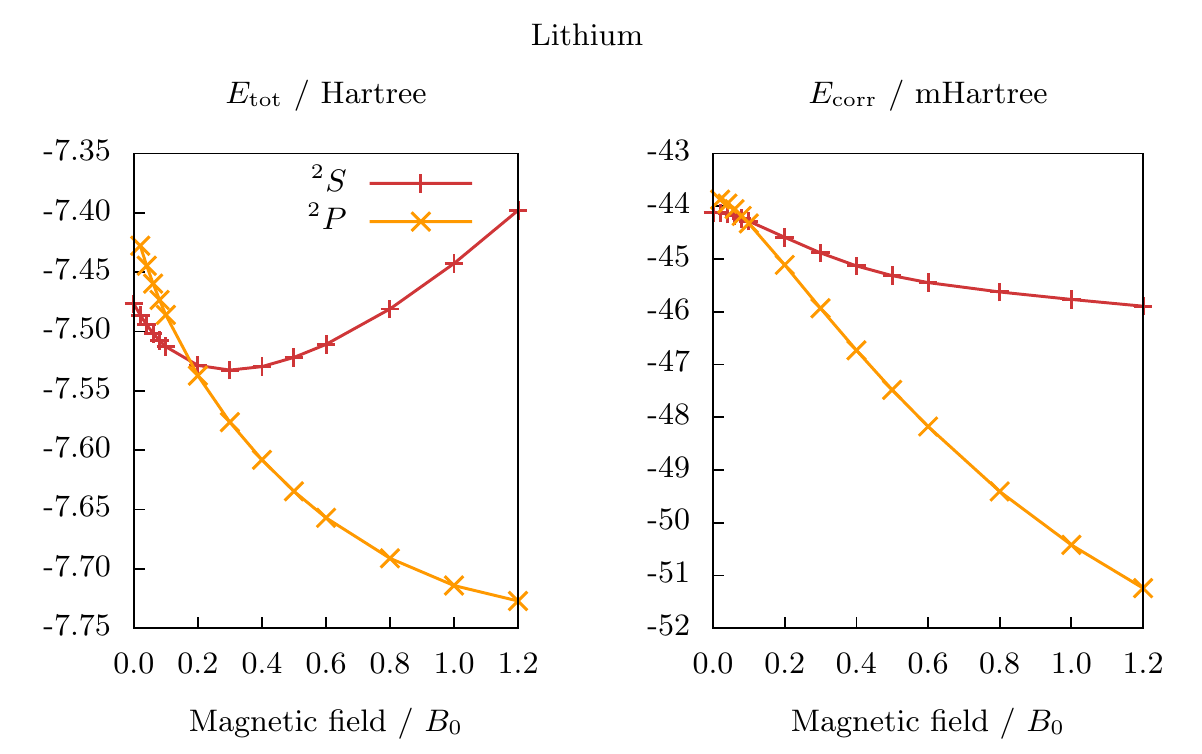}
   \caption{Total and correlation energies of the $^2$S, and $^2$P states of lithium as a function of the magnetic field. 
   Calculated at the CCSD(T) level with the uncontracted aug-cc-pCVQZ basis set.}
   \label{Li}
\end{figure*}

\begin{figure*}
   \includegraphics{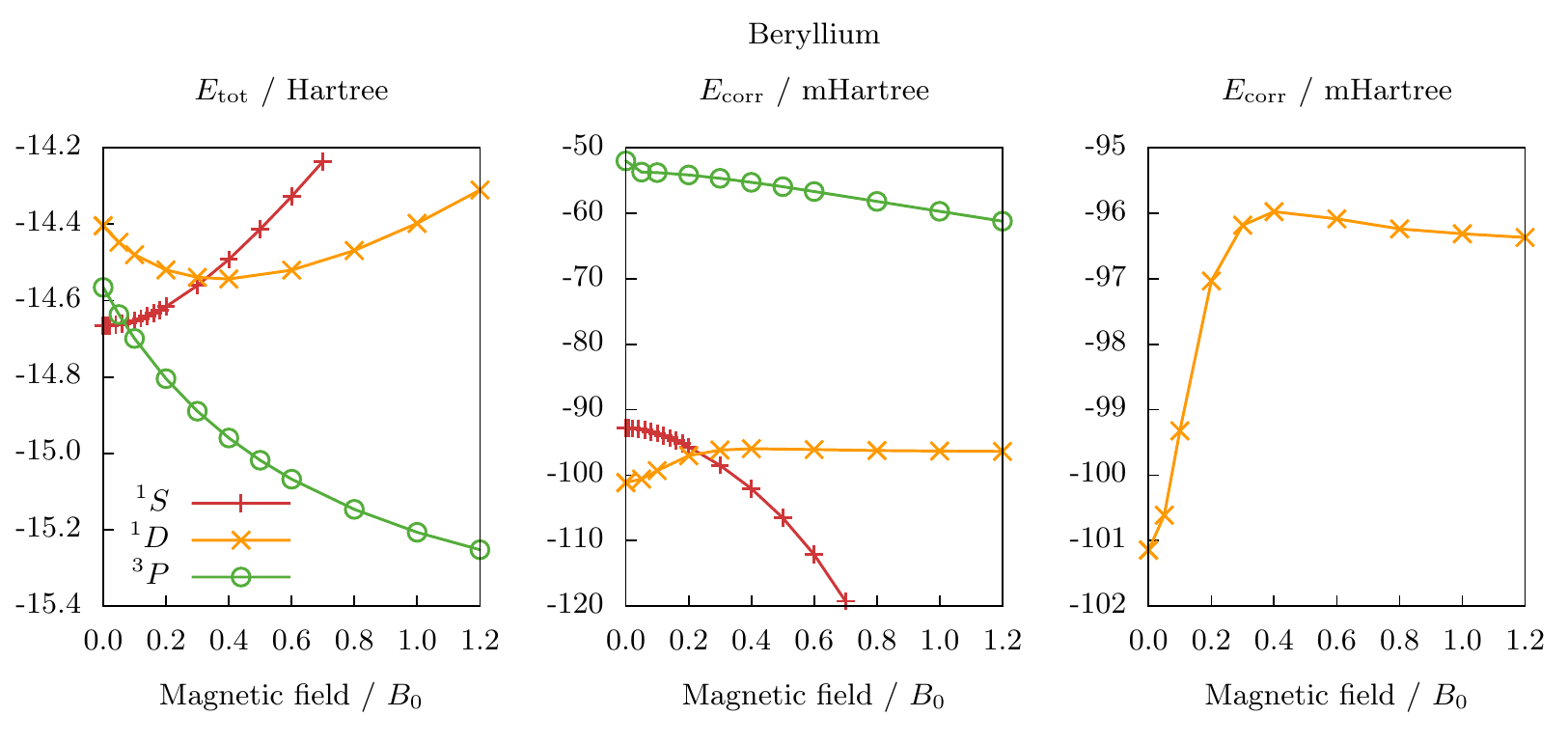}
   \caption{Total and correlation energies of the $^1$S, $^1$D, and $^3$P states of beryllium as a function of the magnetic field.
   Calculated at the CCSD(T) level with the uncontracted aug-cc-pCVQZ basis set.
   }
   \label{Be}
\end{figure*}

\begin{figure*}
   \includegraphics{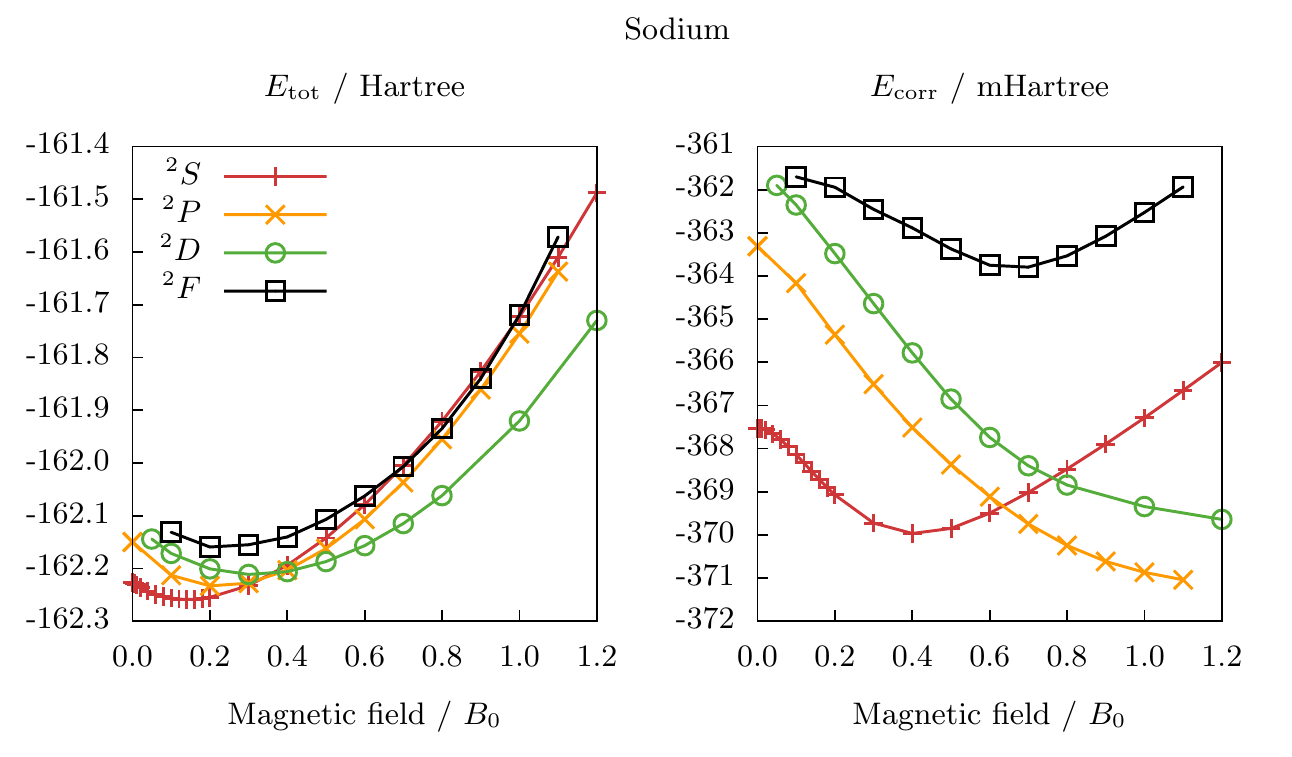}
   \caption{Total and correlation energies of the $^2$S, $^2$P, $^2$D, and $^2$F states of sodium as a function of the magnetic field.
   Calculated at the CCSD(T) level with the uncontracted aug-cc-pCVQZ basis set.}
   \label{Na}
\end{figure*}

The general trend observed so far appears to be that, for states that are more compact at the HF level than at the correlated level, 
correlation energies are reduced in the magnetic field, whereas the opposite is true for states that are more diffuse at the HF level 
than at the correlated level. It is therefore interesting to study lithium, beryllium, and sodium, for which
electron correlation leads, even for the ground state at zero field, to a more compact (rather than diffuse) electronic structure. 
We expect the correlation energies to increase in absolute terms in the presence of the magnetic field.
\\ \\
\textit{Lithium}
\\ \\
For lithium (see Figure \ref{Li}), the $^2$S state is paramagnetically stabilized by the spin-Zeeman term only, see also Ref.\;\onlinecite{Ligroundstate}.
It first goes down in energy but starts to behave diamagnetically at around 0.3\,$B_0$.
At around 0.2\,$B_0$ the $^2$P state becomes more favorable, benefiting from additional stabilization via the orbital-Zeeman term.
As expected from the previous arguments, the correlation energies 
of the $^2$S and $^2$P states, which are of similar magnitude at zero field, both increase in absolute magnitude but
with a steeper increase for the $^2$P state. 
In both cases, the 1s orbital expands in the correlated treatment, leading to a reduced screening experienced 
by the remaining electron in the 2s or 2p orbital.
The steeper rise of electron correlation in the $^2$P state occurs since the 2p orbital shrinks more (see Table \ref{table_spatial}).
\\ \\
\\ \\
\textit{Beryllium}
\\ \\
For the beryllium atom (see Figure \ref{Be}), the diamagnetic $^1$S state is the ground state until around B=0.05\,$B_0$, whereas, 
for higher fields, the para\-mag\-net\-ic $^3$P state is lowest in energy (see also Ref.\,\onlinecite{Beryllium}).
As for lithium, the correlation energy of the corresponding ground state increases in absolute magnitude. 
The correlation-energy curve for the higher $^1$D state is quite peculiar as the correlation energy is first reduced 
but starts to slowly increase in absolute terms around 0.4\,$B_0$.
\\ \\
\textit{Sodium}
\\ \\
For sodium (see Figure \ref{Na}), the three doublet states $^2$S, $^2$P, and $^2$D are quite close in energy up to 0.5\,$B_0$.
Until about 0.3\,$B_0$, the spin-Zeeman stabilized $^2$S state is the ground state, whereafter the  $^2$D state, which is additionally 
stabilized by the orbital-Zeeman term becomes the ground state. 
The $^2$F state is for all field strengths considered here too high in energy to be relevant (at least for the chosen basis set). 
For all considered doublet states, the correlation energy increases in absolute terms with the magnetic field, similarly to the case of lithium. 
However, for the $^2$S state, the correlation energy starts to decrease again in absolute magnitude around 0.4\,$B_0$, 
which most likely can be attributed to the diamagnetic confinement effect.
The reduction of the correlation energy of the $^2$F state for field strengths higher than 0.7\,$B_0$ is most likely an artifact of the basis 
set, as discussed in Section\;\ref{basis_set_chapter}.
\\

Our results for lithium, beryllium, and sodium confirm that, for atoms with only a few valence electrons, 
the correlation energy of the ground state rises at least initially, unlike for systems with many valence electrons such as neon and fluorine.
\subsection{Correlation energies for molecules}
While the study of atoms in strong magnetic fields is already a topic of great relevance---see, for example, 
Refs.\,\onlinecite{ExpevidenceHe,ExpevidenceHe2}---the consideration of molecules adds new aspects to the investigation. 
In comparison with atoms, the main issues are here that, for molecules, there are fewer symmetry constraints and therefore 
more possibilities to respond to the field---see, for example, Refs.\,\onlinecite{zitatschmelcher,KaiScience,HX,H2,H2numbers}.
Also, there is additional flexibility as bond distances and angles depend on the magnetic field, making it necessary to investigate the optimal bond distances and potential energy surfaces as a function of the field.
The symmetry issue deals here mostly with the question whether the angular momentum is a good quantum number or not. 
While this is always true for atoms, it does not necessarily hold for molecules, which opens opportunities for paramagnetic stabilization and additional electron-correlation effects.

\begin{figure*}
   \includegraphics{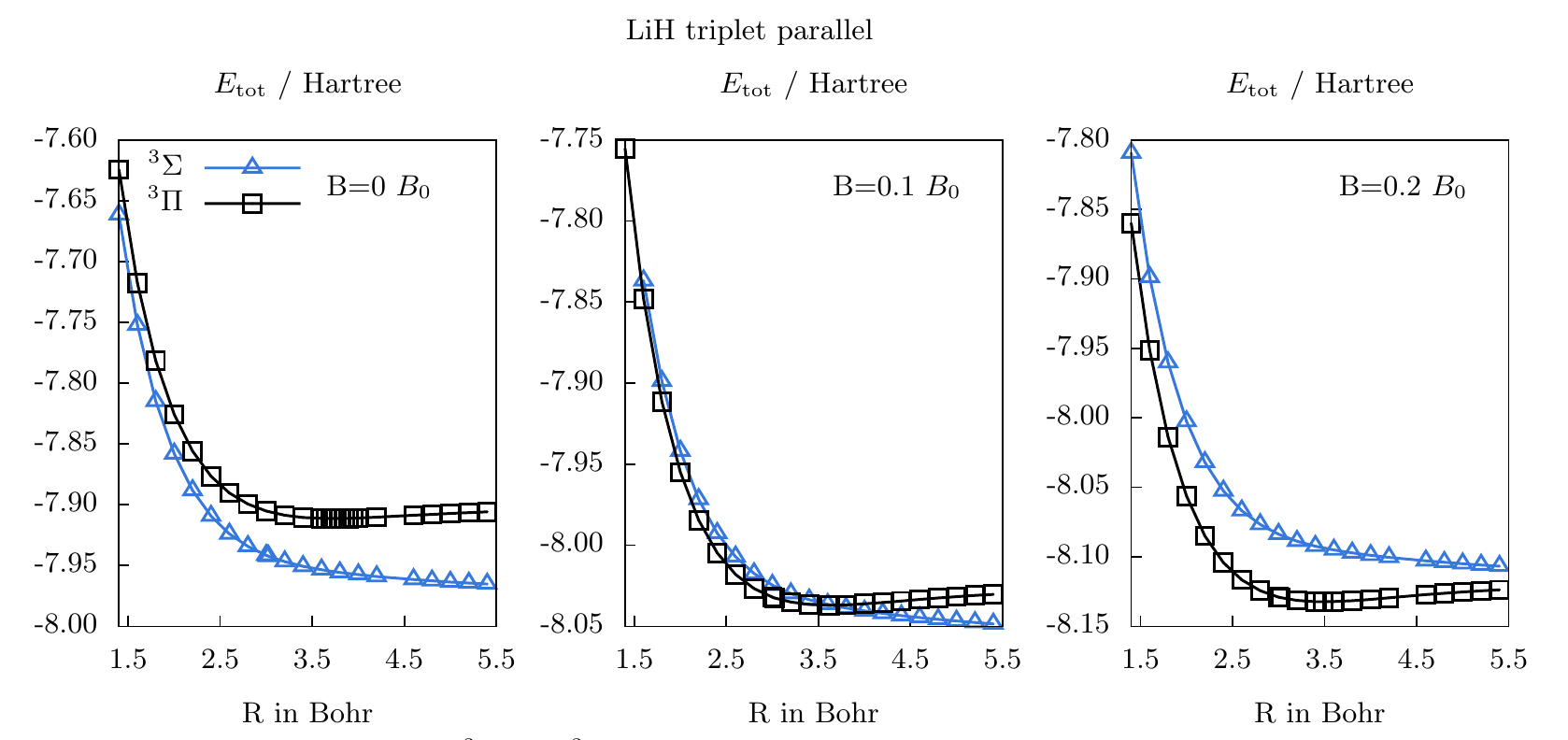}
   \caption{Potential-energy curves for the  $^3\Pi$ and $^3\Sigma$ state of LiH.
   Left: zero-field, middle: $\text{B}=0.1\,B_0$ (parallel orientation), right: $\text{B}=0.2\,B_0$ (parallel orientation). Calculated at the CCSD(T) level with the uncontracted aug-cc-pVTZ basis set.}
   \label{LiH_Triplet_para_pi_vs_sigma}
\end{figure*}

\begin{figure*}
   \includegraphics[width=0.7\textwidth,trim=0cm 0cm 0cm 0.1cm,clip=true]{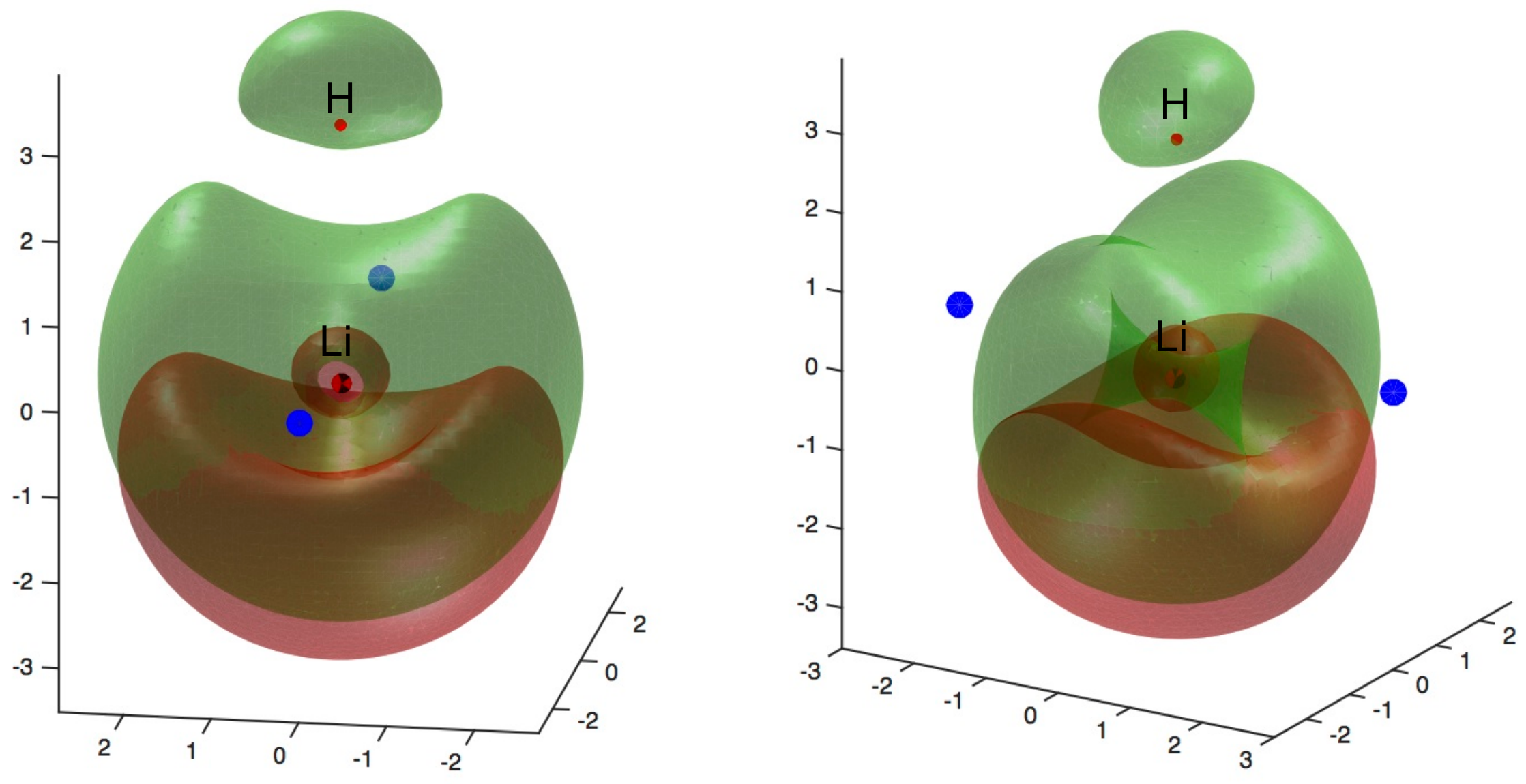}
   \caption{HOMO (absolute value of the complex orbital) of the triplet state of LiH viewed from two different angles in the perpendicular 
orientation at $\text{B}=0.2\,B_0$ (red) and $\text{B}=0.6\,B_0$ (green), isosurface cut at 0.07 a.u. The blue dots indicate the direction of the field.}
   \label{HOMO}
\end{figure*}

We focus in the following on the singlet and triplet states of LiH, discussing the correlation energies for orientations 
with the field parallel and perpendicular to the molecular axis. 
Furthermore, we investigate correlation energies in a magnetic field at the equilibrium distance in that field as well as 
study the dependence of the correlation energy for a fixed geometry on the magnetic field. 
\\ 
\\
\begin{figure*}
   \includegraphics{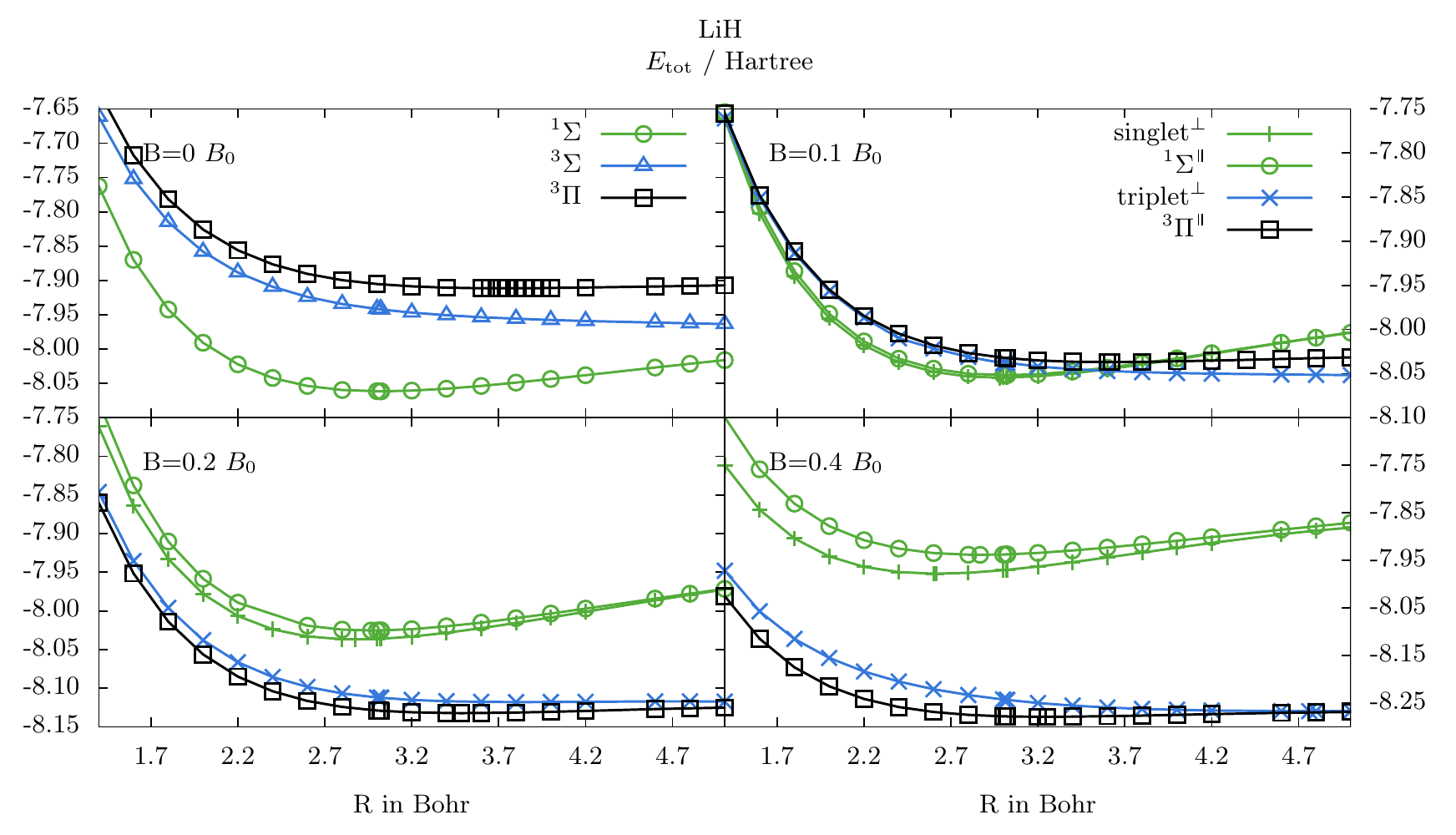}
   \caption{Potential-energy curves for the lowest singlet and triplet states of LiH in a magnetic field of 0, 0.1, 0.2, and 0.4\,$B_0$. For non-zero fields, the curves for the perpendicular and parallel orientation of the field with respect to the molecular axis are shown. Calculated at the CCSD(T) level with the uncontracted aug-cc-pVTZ basis set.}
   \label{LiHgeneral}
\end{figure*}
In the field-free limit, the $^1\Sigma$ state of LiH is the ground state. 
As the field strength increases gradually, the energy of the singlet state increases, 
while the energy of the lowest-lying triplet state decreases due to paramagnetic stabilization (see Figure \ref{LiHgeneral}). 
At a field strength of around 0.2\,$B_0$, the triplet state becomes the ground state. 
At higher fields, the singlet and triplet states move even further apart and the energy splitting between the parallel and 
perpendicular orientations becomes apparent. 
In the parallel orientation, at a field strength of around 0.1\,$B_0$, 
the weakly bound $^3\Pi$ state becomes lower in energy than the unbound $^3\Sigma$ state, 
as the occupation of the $\pi$ orbital with $m_l=-1$ provides additional paramagnetic stabilization (see Figure \ref{LiH_Triplet_para_pi_vs_sigma}). 
For that reason, we plot for all non-zero fields in Fig.\,\ref{LiHgeneral} the $^3\Pi$ state in the parallel orientation.
However, in the perpendicular orientation, 
the $^3\Sigma$ and $^3\Pi$ distinction is no longer meaningful. As the magnetic field is turned on, the symmetry is broken and the initial
$\Sigma$ state mixes with $\Pi$ states---see Figure\;\ref{HOMO}, which demonstrates the increased $\pi$ character of the highest occupied molecular orbital (HOMO) for $\text{B}=0.6\,B_0$).

\begin{figure}
   \includegraphics{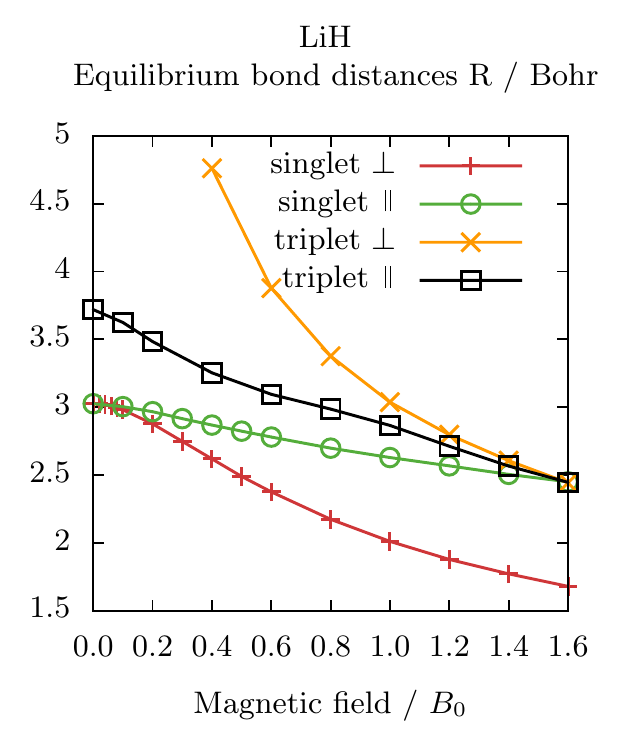}
   \caption{Equilibrium bond distances of the lowest singlet and triplet states of LiH as a function of the magnetic field. Calculated at the CCSD(T) level with the uncontracted aug-cc-pVTZ basis. }
   \label{LiH_Req_vs_B}
\end{figure}

For all states considered, the equilibrium bond distances shorten in the magnetic field (see Figure \ref{LiH_Req_vs_B}),
as the orbitals become more compact.
\subsubsection{Correlation energies at equilibrium bond distances}
For the singlet states in both the perpendicular and the parallel orientations, 
the correlation energy in the field initially decreases in absolute magnitude due to confinement (see Figure\,\ref{LiH_singlet_corr_Rfixed_and_Req}).
This decrease is stronger in the parallel orientation than in the perpendicular case. 
At around $0.08\,B_0$ in the perpendicular case and $0.2\,B_0$ in the parallel case, the correlation energies start to rise.
At $1\,B_0$, the increase is 2.2\% in the parallel orientation and as much as 11\% in the perpendicular case relative to 
the zero-field correlation energies. 

This increase may be understood by the observation that, even though we are considering equilibrium distances, the bonding 
situation does not remain the same. 
When the field is turned on, the bond distances shorten, bringing the electrons closer together and increasing electron correlation. 
As the bond distance is substantially shorter in the perpendicular case (see Figure \ref{LiH_Req_vs_B}), 
the increase in correlation energy is more pronounced in that orientation.

For the triplet state (see Figure \ref{LiH_triplet_corr_all}), as for the atoms investigated in this study, 
an increase in correlation energy is observed in both orientations.
Note that, in the parallel orientation, we refer to the weakly bound $^3\Pi$ state, 
whereas, for the lowest state in the perpendicular orientation, the onset of bonding occurs only around 0.4\,$B_0$.
Initially, the correlation energies for the two states are therefore quite different but they become more similar 
with increasing field strength as the state in the perpendicular orientation adopts more $\pi$ character (see next section).
\subsubsection{Correlation energies at a fixed geometry}
\begin{figure}
   \includegraphics{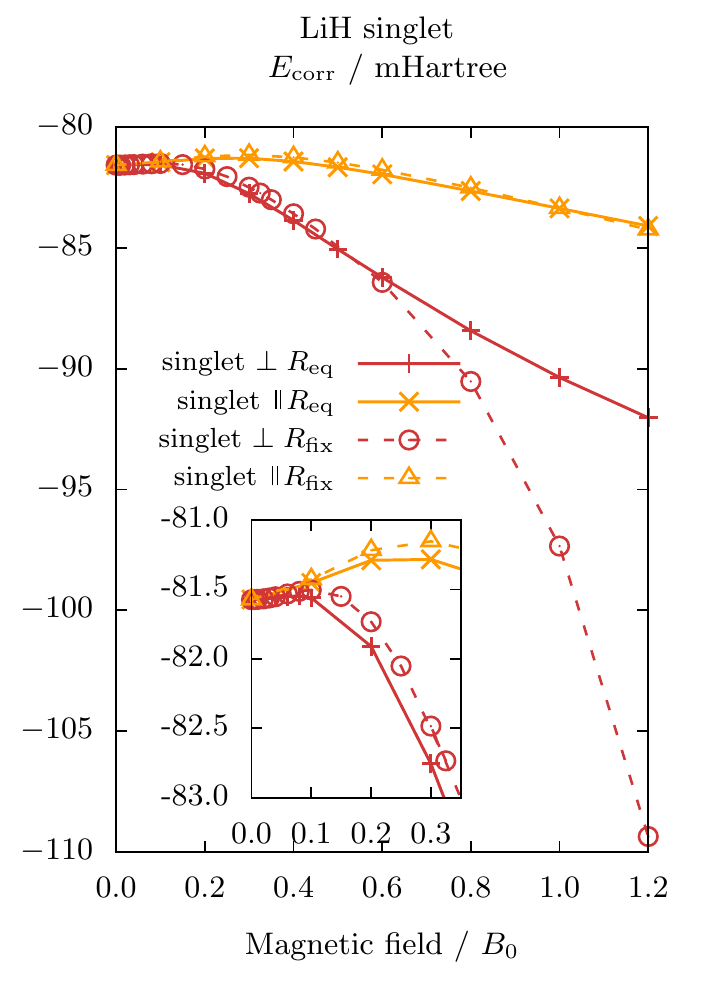}
   \caption{Correlation energy of the singlet state of LiH in parallel and perpendicular orientation at equilibrium and fixed bond distance ($R$=3.0248\;bohr) as a function of the magnetic field.
   Calculated at the CCSD(T) level with the uncontracted aug-cc-pCVQZ basis set.
   }
   \label{LiH_singlet_corr_Rfixed_and_Req}
\end{figure}

For discussing the field dependence of the correlation energies at a fixed geometry, we consider a distance of 3.0248\;bohr, 
which corresponds to the calculated equilibrium distance of the $^1\Sigma$ state in the field-free case (CCSD(T), uncontracted aug-cc-pVTZ basis). 
As seen from Figure\,\ref{LiH_singlet_corr_Rfixed_and_Req}, for the singlet state in the perpendicular and parallel orientations, 
a similar picture as for the correlation energies at the equilibrium distances emerges. 
There is first a reduction in electron correlation in absolute terms due to the confinement (up until 0.1\;$B_0$ for the perpendicular case and 
0.3\;$B_0$ for the parallel case), followed by a rise in the correlation energy. 
Fixing the distance means moving towards the dissociation limit 
(thereby increasing static correlation due to the dissociation into  Li ($^2$P) and H ($^2$S)) 
as the equilibrium bond lengths decrease considerably with increasing field. 
This increase in correlation energy is again much stronger in the perpendicular case, due to the more pronounced shortening of the equilibrium bond distance (see Figure \ref{LiH_Req_vs_B}). 
For the same reason, the correlation-energy curves for the fixed and optimized distances are very similar (see Figure\,\ref{LiH_singlet_corr_Rfixed_and_Req}) in the perpendicular orientation, as long as the distances do not differ too much.

\begin{figure}
   \includegraphics{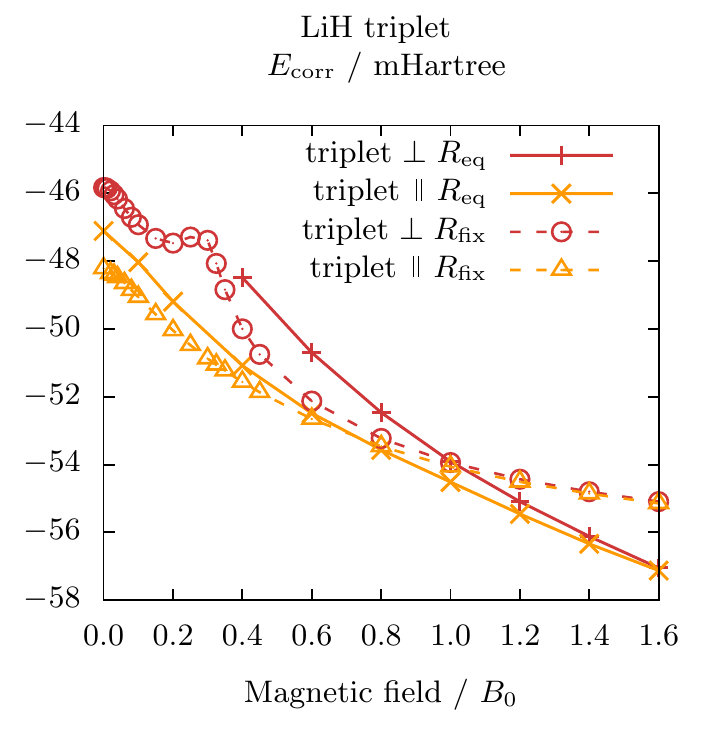}
   \caption{Correlation energy for the triplet state of LiH in parallel and perpendicular orientation, both at equilibrium and fixed bond distance (R=3.0248 bohr) as a function of the magnetic field.
   Calculated at the CCSD(T) level with the uncontracted aug-cc-pCVQZ basis set.
   }
   \label{LiH_triplet_corr_all}
\end{figure}

For the triplet state in the parallel orientation, we observe a general raising of the correlation energy 
with the magnetic field but with different slopes at either side of $\mathrm{B}=0.6\,B_0$ (see Figure\;\ref{LiH_triplet_corr_all}).
At this field strength, the fixed geometry is closest to the actual (field-dependent) equilibrium geometry of the system. 
At lower field strengths, the equilibrium distance is longer and the results correspond to a system compressed towards the 
united-atom limit; conversely, for higher field strengths, the results correspond to a stretched system.
While going towards the united-atom limit raises the correlation energy for a given field strength, 
moving towards the dissociation limit lowers the correlation of the triplet state.
Therefore, when correlation energies for fixed and 
optimized distances are compared (see Figure \ref{LiH_triplet_corr_all}), 
the fixed-geometry curve lies beneath the corresponding equilibrium-geometry 
for field strengths up to 0.6\,$B_0$; for stronger fields, the situation is reversed. 

\begin{figure}
   \includegraphics{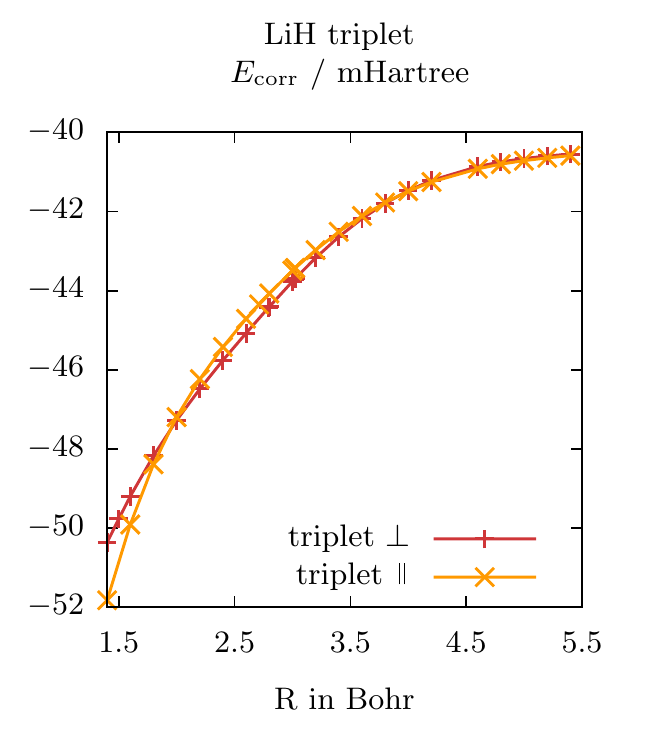}
   \caption{Correlation energy for the triplet states in perpendicular and parallel orientation as a function of the bond distance at a field strength of 1.2 $B_0$.
   Calculated at the CCSD(T) level with the uncontracted aug-cc-pVTZ basis set.
   }
   \label{LiH_triplet_corr_vs_RatB1_2}
\end{figure}

For the triplet state in the perpendicular orientation, the dependence of the correlation energy on the magnetic field is more complicated:
the correlation energy first increases in absolute terms with increasing field until 0.2\,$B_0$, then decreases until about $0.3\,B_0$ 
and rises again for higher field strengths.
The reason for this puzzling behavior is that the HOMO changes character (see Figure \ref{HOMO}), 
from a deformed sp-hybrid orbital to a $\pi_{-1}$ orbital.
As the bond distances for larger fields converge to more or less the same values in the parallel and the perpendicular case, 
the correlation energies are also becoming similar. 
The correlation energy for the triplet state is reduced when moving towards the dissociation limit at a fixed field strength 
(Li($^2$P) and H($^2$S) coupled to a triplet) as this limit can be well described with a single determinant and does not 
introduce static correlation (see Figure\,\ref{LiH_triplet_corr_vs_RatB1_2}). 
The fixed bond distance of 3.0248\,bohr is for higher fields significantly longer than the equilibrium distance, 
explaining why those correlation energies are smaller in absolute magnitude than those obtained at the equilibrium geometry. 

Our discussion for the various LiH states reveals that the dependence of the correlation energy in molecules on the magnetic 
field is determined by a variety of effects. 
Besides those present in atoms such as the confinement, we have to consider also symmetry lowering leading to paramagnetic stabilization, 
changes in the equilibrium geometry, and possible changes in the dissociation limits with the appearance of static correlation. 
The investigation of more molecules is necessary for a more complete understanding of these effects and their interplay.
\\
\subsection{\red{Basis-set dependence and comparison with literature values}}
\label{basis_set_chapter}
\begin{figure*}
   \includegraphics{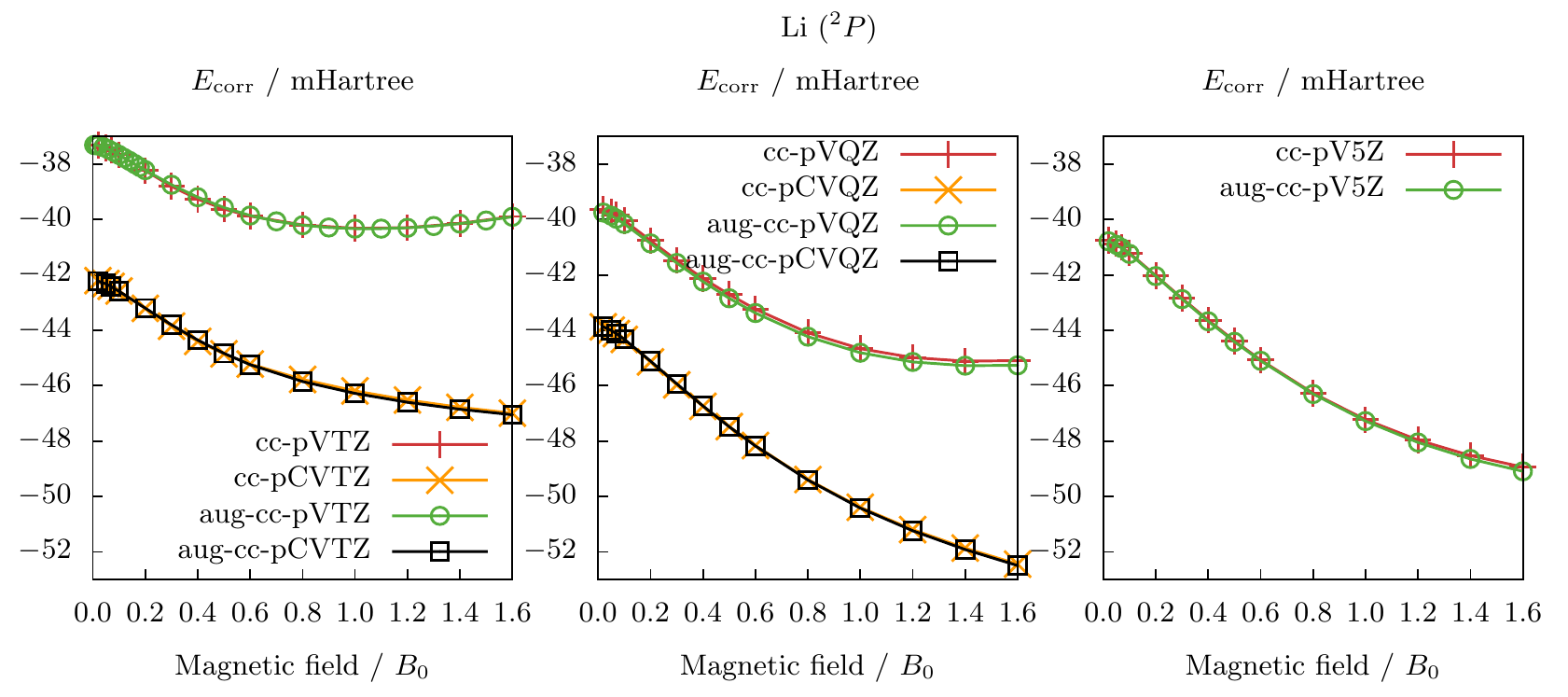}
   \caption{Basis-set convergence for the correlation energy of the lithium $^2$P state in CCSD(T) calculations with finite magnetic fields. All basis sets were uncontracted.}
   \label{Li_basis}
\end{figure*}
An important issue in electron-correlated calculations is always basis-set convergence.\cite{lilabibel}
In the present case, we must deal with the additional difficulty of adequately describing the 
magnetic-field effects and their interplay with electron correlation.
Gauge-origin independence is here of no concern as we use GIAOs.
However, the magnetic field leads to a deformation and compression of the orbitals \cite{KaiScience} that need to be described with the chosen basis set.
This problem may be best dealt with by using anisotropic Gaussians.\cite{aniso1,aniso2} Here, we examine the performance of standard uncontracted Gaussian basis sets for calculations on atoms and molecules in strong magnetic fields. 

For the $^2$P state of lithium, we compare in Figure\;\ref{Li_basis} the correlation energies obtained with the uncontracted cc-pVXZ and aug-cc-pVXZ (X=T,Q,5) as well as the cc-pCVQZ and aug-cc-pCVXZ (X=T,Q) basis sets. 
All these basis sets raise electron correlation up until around 1\,$B_0$.
There is essentially no difference in the result for the correlation energies obtained with basis sets with and without augmentation by diffuse functions.
For the smallest basis set considered here---namely, cc-pVTZ and aug-cc-pVTZ---a reduction in the correlation energy is observed for higher fields that is not seen for the larger basis sets, strongly indicating that this reduction is a basis-set artifact.
Both, increasing the cardinal number of the basis set and adding core-correlating functions leads to a raise in correlation energy.
More importantly, the artificial reduction of the correlation energy for higher fields is lessened and its onset is shifted towards higher magnetic fields.

These findings indicate that an unexpected reduction of correlation energy in higher magnetic fields should be viewed with some caution.
We also note that the inclusion of core-polarization functions does not change the qualitative dependence of the correlation energy 
on the magnetic field, even though the magnitude of the correlation energy is significantly enhanced and the onset of 
the unphysical behavior occurs for slightly higher fields. 
We conclude that our results for the correlation energy can be trusted only up until roughly 0.8\,$B_0$ for triple-zeta basis sets, 
while the results obtained using the uncontracted aug-cc-pCVQZ basis set seem to be reasonable until 1.4\,$B_0$.

\begin{figure*}
   \includegraphics{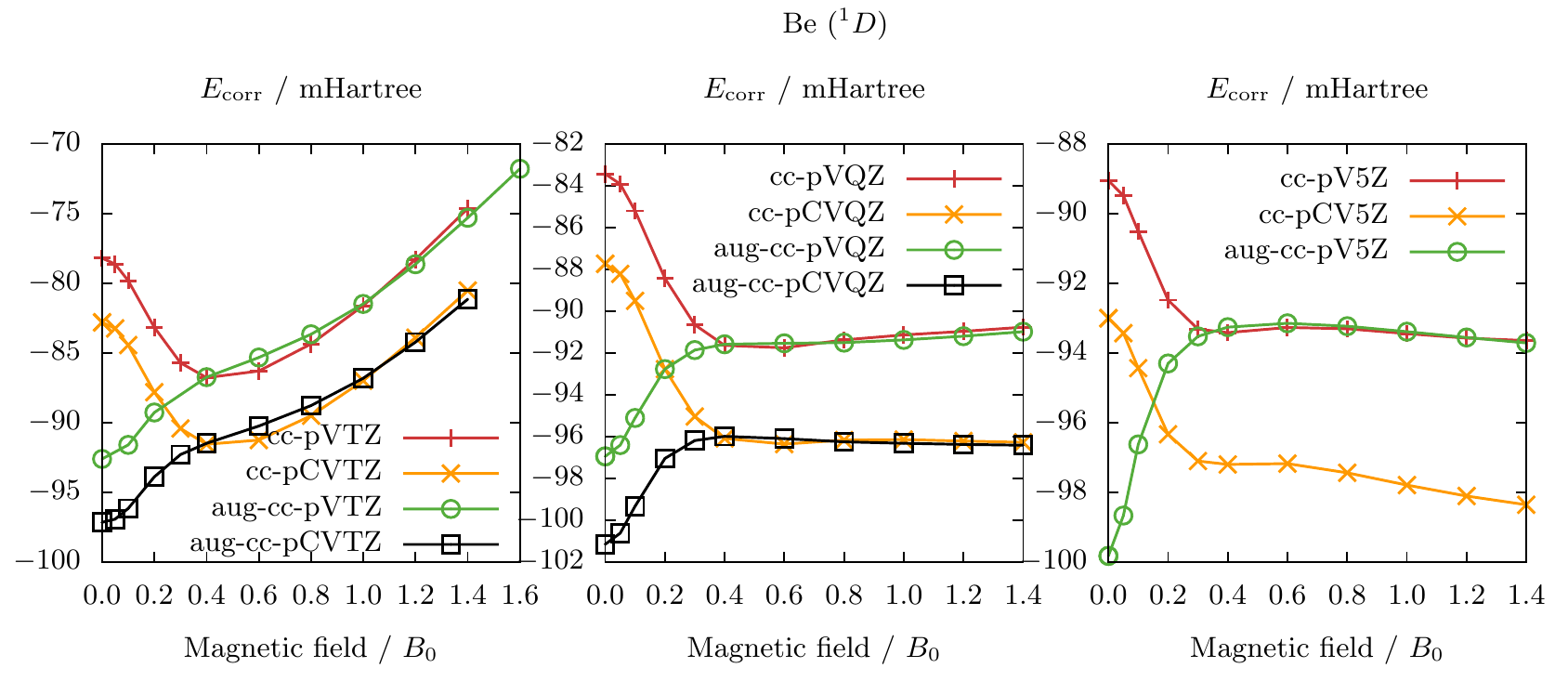}
   \caption{Basis-set convergence for the correlation energy of the $^1$D state of beryllium in CCSD(T) calculations with finite magnetic fields.
   All basis sets were uncontracted.}
   \label{Be_basis}
\end{figure*}

To assess the quality of the description for states of higher angular momentum, we have here investigated the $^1$D  state of beryllium,
comparing results obtained using the uncontracted cc-pVXZ, cc-pCVXZ, aug-cc-pVXZ (X=T,Q,5), and aug-cc-pCVXZ (X=T,Q) basis sets (see Figure \ref{Be_basis}).
Interestingly, for all cardinal numbers, we find that, up to 0.4\,$B_0$, 
augmentation by diffuse functions reverses the trend seen in the basis-set dependence of the correlation energy. 
Whereas  sets without augmentation initially predict a rise in the correlation energy, those with diffuse functions predict a reduction. 

A closer analysis indicates that it is not the addition of diffuse p functions that is crucial but rather the augmentation by additional d functions. 
The dominant double excitation for this state is from 2p$_{-1}$2p$_{-1}$ to 3s3d$_{-2}$. 
As the diffuse d functions lower the energy of the 3d$_{-2}$-orbital significantly (from 0.3785\,Hartree for the cc-pCVQZ basis to 
0.1424\,Hartree for the aug-cc-pCVQZ basis in the field-free case), the amplitude for the dominant excitation increases (from 0.20 to 0.38), leading also to an enhanced correlation energy.
It should be noted, however, that the $^1$D state of beryllium is a multi-reference case not well described with CCSD(T) theory.

For higher fields, the results for augmented and non-augmented sets show similar behavior and here the contraction of the orbitals in the magnetic field seems to be the dominating effect. 
Furthermore, adding tight core functions leads generally to a more or less constant shift in the correlation energy, as already discussed for the $^2$P state of lithium. 
The sets of triple-zeta quality again show an unphysical reduction in correlation energy for higher fields, which is not observed for basis sets with higher cardinal numbers.
In fact, results obtained with basis sets of quintuple zeta quality indicate a slight rise of correlation energy for higher fields.
Obviously, states with high angular momentum have more severe basis-set requirements. 
This is already the case in the zero-field limit but is even more pronounced when moving to high magnetic fields.
A detailed analysis reveals that, in the triple-zeta basis sets, the s and p block is too small to correctly account for the contraction of the orbitals.
\\ \\
\begin{figure*}
   \includegraphics{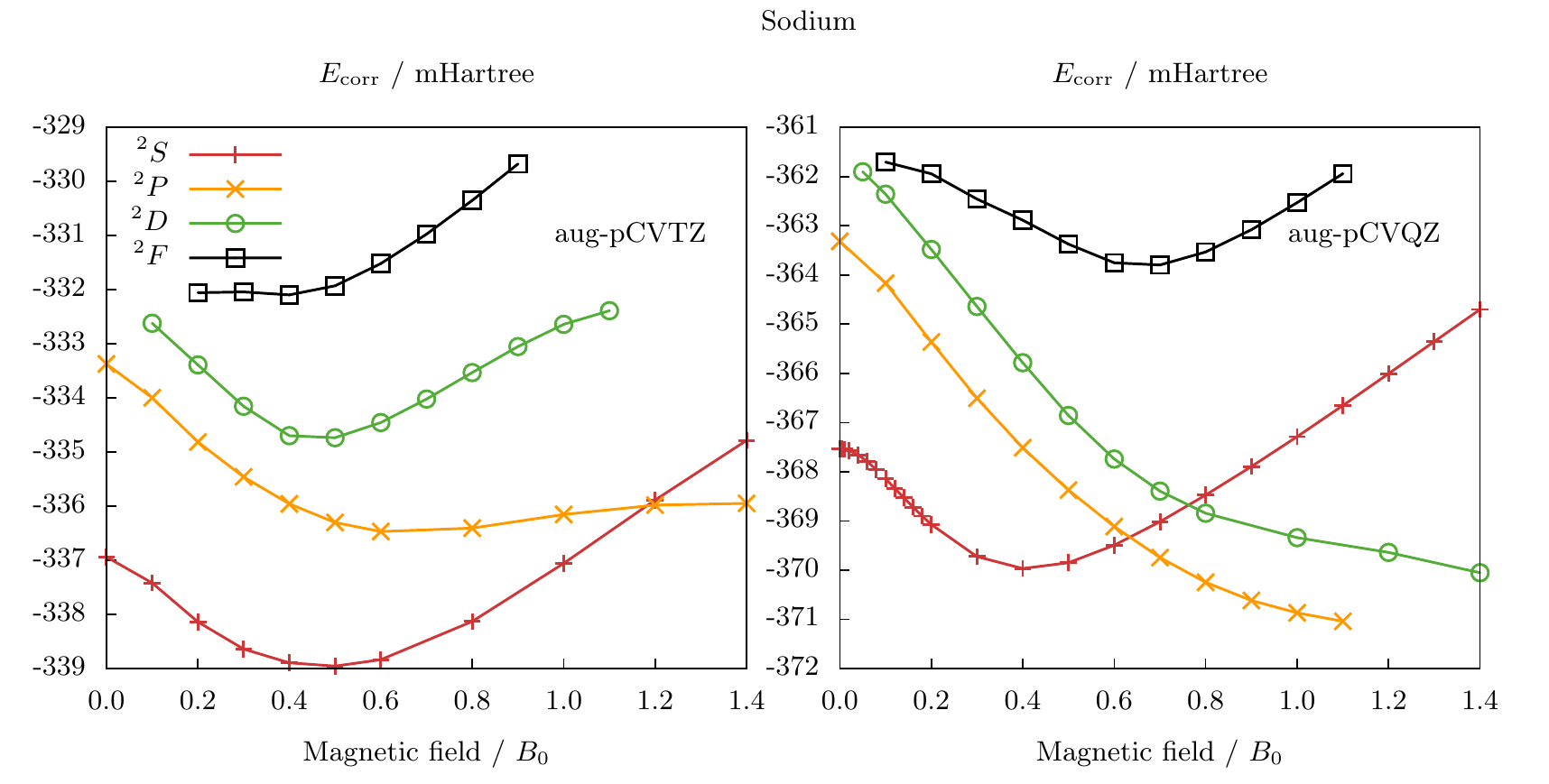}
   \caption{Basis-set convergence for the correlation energy of the $^2$S, $^2$P, $^2$D, and $^2$F states of sodium in CCSD(T) calculations with finite magnetic fields.
   All basis sets were uncontracted.}
   \label{Na_basis}
\end{figure*}
For sodium, we compare in Figure \ref{Na_basis} the basis-set dependence of the correlation energy of the $^2$S, $^2$P, $^2$D, and $^2$F states computed with the aug-cc-pCVXZ (X=T,Q) basis sets. 
While the $^2$S state is already reasonably well described with the aug-cc-pVTZ basis set, we can clearly see that the aug-cc-pCVTZ description deteriorates for states with higher angular momentum and that the unphysical "bending up" of the correlation energy curve occurs at even lower field strengths. 
Even the qualitative form of the correlation energy curve can only be trusted until 0.6\,$B_0$ for the $^2$P state and 0.4\,$B_0$ for the $^2$D state, respectively, with the aug-cc-pCVTZ basis. 
For the $^2$F state, it is not even clear whether it is sufficiently well described with the aug-cc-pCVQZ basis set.
From our findings, we conclude that standard basis sets are problematic both when considering atoms or molecules in higher magnetic fields as well as  when describing states of higher angular momentum and, in particular, when trying to describe both. 
While those states are often of no interest in the field-free case, they are stabilized by paramagnetic effects and are therefore of relevance in (strong) magnetic fields. 

\begin{figure}
   \includegraphics{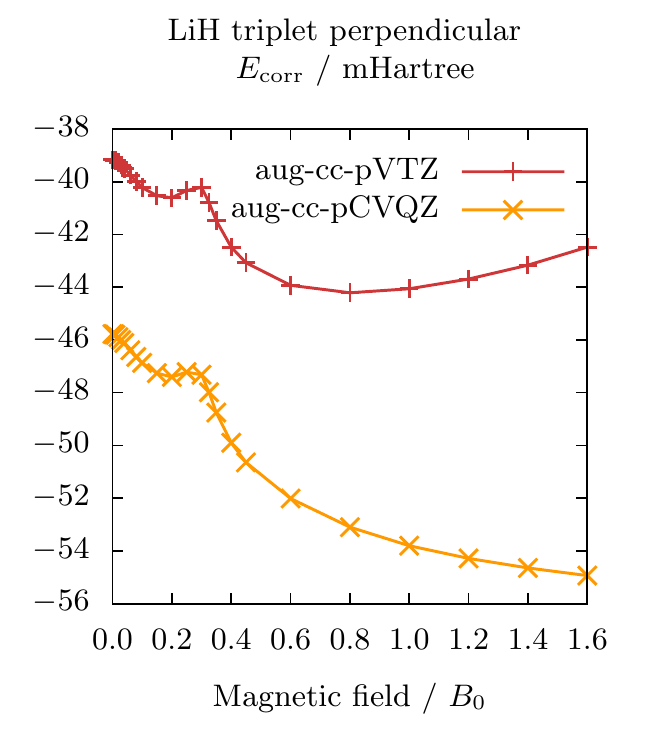}
   \caption{Basis-set convergence for the correlation energy of the triplet state of LiH in perpendicular orientation at a fixed bond distance of R=3.0248 Bohr, calculated at the CCSD(T) level with finite magnetic fields. All basis sets were uncontracted.}
   \label{LiH_triplet_perp_basis}
\end{figure}

For the triplet state of LiH in a perpendicular orientation, we again find an artificial reduction of the correlation energy for high field strengths, 
as can be seen when comparing the curves for the unconctracted aug-pVTZ and aug-pCVQZ basis sets in Figure  \ref{LiH_triplet_perp_basis}.
Here, we compare results obtained with the uncontracted aug-cc-pVTZ and the uncontracted aug-cc-pCVQZ basis set. 
While the curves are similar until 0.6\,$B_0$, for higher fields the correlation energy decreases in absolute terms for the aug-cc-pVTZ basis but actually rises in the case of the aug-cc-pCVQZ basis set.

\red{
Finally, in Fig. \ref{litcomp}, we compare our results obtained using uncontracted isotropic Gaussian basis sets with FCI values from the literature\cite{HeliumFCI,HeliumFCIp,LithiumFCI,BerylliumCI,H2numbers,H2numbersPi,Schmelchercorr} obtained using anisotropic Gaussians as basis sets. 
\begin{figure}
   \includegraphics{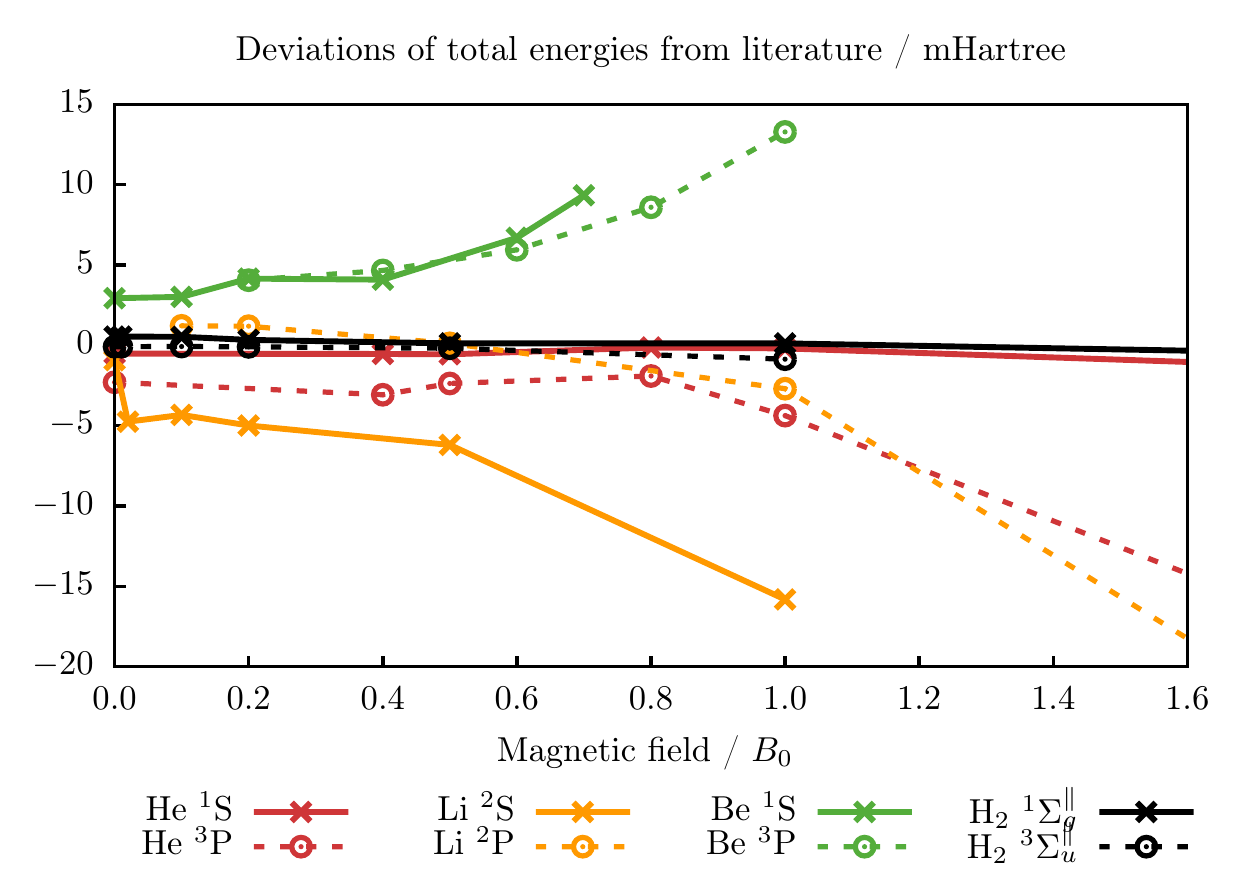}
   \caption{Deviations of CCSD(T) total energies (computed using the doubly augmented uncontracted d-aug-cc-pCVZ in the case of helium and the uncontracted aug-cc-pCVQZ basis set in all other cases) from literature data computed using anisotropic Gaussian basis sets as a function of the magnetic field.}
   \label{litcomp}
\end{figure}
A detailed comparison is found in Table XXIV of the supplementary information. \footnote{ See supplementary material for additional details.
}
Differences in the wave-function method are here of minor importance as the number of electrons is so small that our CCSD(T) approach is identical (for two electrons) or very similar (for three or four electrons) to the FCI method. 
A comparison therefore reveals the performance of uncontracted standard basis sets in contrast to anisotropic ones.
For example, the errors in correlation energy are of the order of only 10$^{-6}$ and 10$^{-5}$ Hartree for the Li $^2$S and Be $^1$S states at zero field at the CCSD(T) level, respectively, while the basis-set effects are at least one order of magnitude larger. 
\\
Our goal is to treat various systems using a reasonably large basis set rather than aiming for the most accurate result obtainable for each system within our method. 
Therefore, the uncontracted aug-cc-pCVQZ basis-set is used in all cases except for helium, for which an uncontracted doubly augmented d-aug-cc-pVQZ basis set is used to provide p functions suitable for describing the $^3$P state.  
We find good agreement with data from the literature for the considered states, with basis-set errors of about $10^{-3}$ Hartree or less for field strengths smaller than 1 $B_0$.
As expected, P states are slightly less well described than S states, with deviations increasing towards higher field strengths.
The discrepancies in the total energies for the $^1$S state of helium, for example, are of the order of $10^{-4}$ Hartree up to a field strength of 1 $B_0$, while they are of the order of $10^{-3}$ Hartree for the $^3$P state. 
\\
For lithium, the differences are less pronounced between the $^1$S and $^2$P states, being in both cases of the order of $10^{-3}$ Hartree for fields lower than 1 $B_0$.
Curiously, the energies of the $^2$P state obtained in our calculations are lower than the literature values\cite{LithiumFCI} for field strengths lower than $1 B_0$, suggesting that in this particular case the basis set chosen in the literature may have been too small. 
\\
For beryllium, our calculated energies are consistently lower than the literature values\cite{BerylliumCI}, which can be attributed to the fact the latter have been obtained using a frozen-core approximation neglecting the correlation between the core and the valence electrons (estimated by the authors to be of the order of $10^{-3}$ Hartree) whereas in our calculations all electrons are correlated with respect to each other.
In this case, the missing coupling between core- and valence correlation is more important than the use of anisotropic basis functions.\\
In conclusion, the use of standard basis sets of aug-cc-pCVQZ quality appears justified for field-strengths up to around 1 $B_0$.
}
\subsection{Binding energies for molecules}
In this section, we report on the binding energies for the singlet states of LiH and triangular He$_3$ in the perpendicular direction of the magnetic field and, in particular, consider the contributions due to electron correlation.

For the binding energy of LiH as a function of the field strength, we find a peak at 0.2\,$B_0$ (see Figure \ref{LiH_singlet_perp_bonden}) due to the fact that the dissociation limit changes:
\begin{figure*}
   \includegraphics{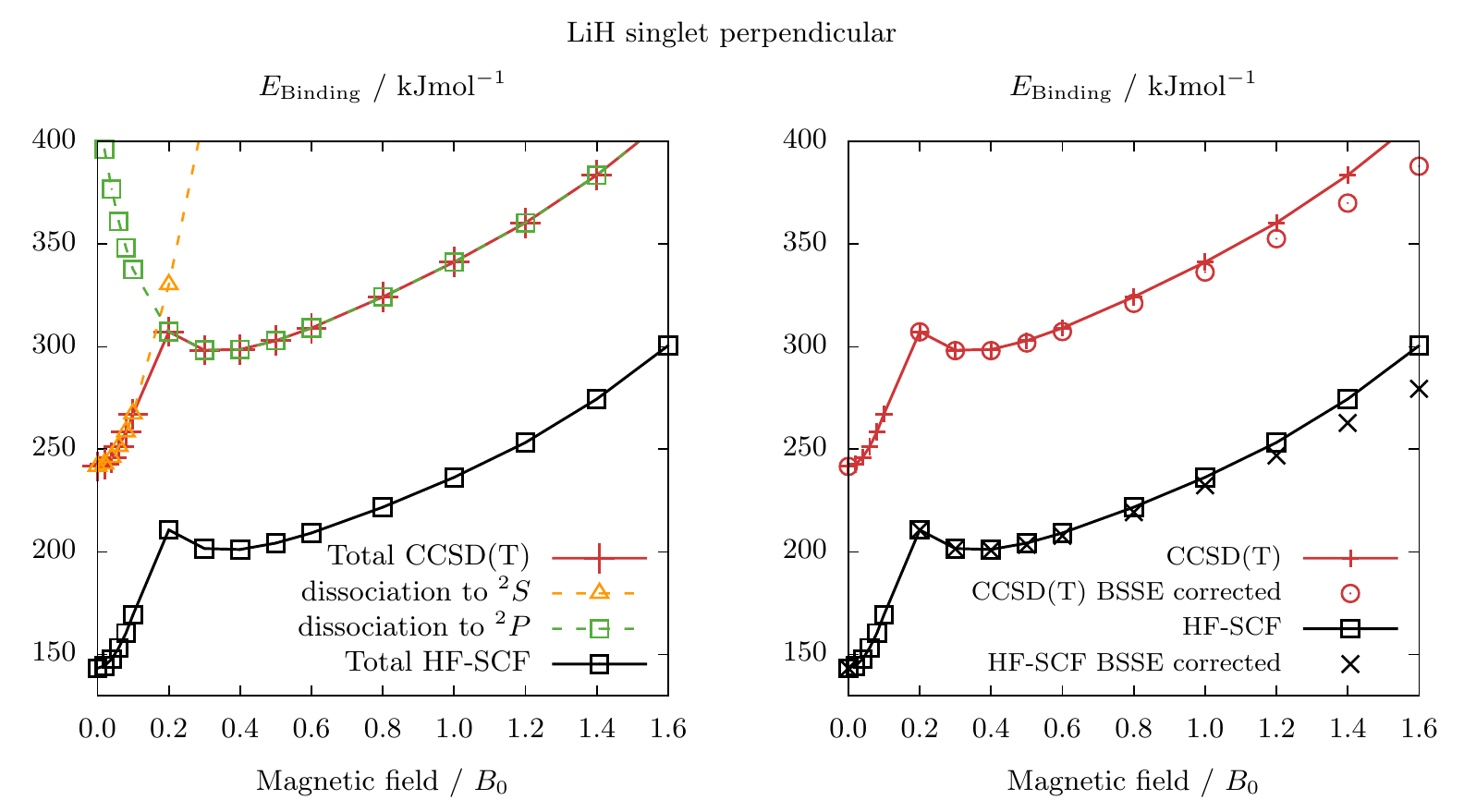}
   \caption{Binding energy for the singlet state of LiH in perpendicular orientation as a function of the magnetic field. 
   Left: CCSD(T) (red) and HF-SCF (black) binding energy as a function of the magnetic field. 
   For CCSD(T) also the binding-energy curves for the dissociation to the $^2$S state (orange) and to the $^2$P state of lithium (green) are plotted. 
   Right: Comparison of binding-energy curves with (dotted lines) and without (full lines) BSSE correction for CCSD(T) (red) and HF-SCF (black).
   Calculations were performed using the uncontracted aug-cc-pCVQZ basis set.
}
   \label{LiH_singlet_perp_bonden}
\end{figure*}
Initially, and for fields smaller than about 0.2\,$B_0$, the dissociation of LiH leads to the $^2$S state of Li, 
while for higher fields the $^2$P state with $m_l=-1$ is lower in energy. 
For both dissociation limits, the binding energy increases with the field.
Results have been obtained with and without basis-set superposition error (BSSE) correction.\cite{counterpoise,BSSE}
Clearly, for magnetic fields larger than 1\,$B_0$, the BSSE becomes significantly larger than 5\,kJ/mol (see Table \ref{table_BSSE}), indicating the inadequacy of the basis set used for higher magnetic fields. 
The relative BSSE is less than 5$\%$ for fields up to 1.4\,$B_0$.
\begin{table}[h,t,b,p]
\begin{footnotesize}
\caption{
Binding energies and BSSE for LiH and He$_3$ calculated at the CCSD(T) level with the uncontracted aug-cc-pCVQZ basis set. Geometries were obtained at the CCSD(T) level with the uncontracted aug-cc-pVTZ basis set.
}
\label{table_BSSE}
\begin{tabular*}{\linewidth}{l@{\extracolsep\fill}  c c c}
\hline
\hline
B$/B_0$ & Binding energy/kJmol$^{-1}$ & BSSE/kJmol$^{-1}$ & BSSE/$\%$ \\
\hline
\multicolumn{4}{c}{LiH}\\
0.0     &       241.67  &       0.02    &       0.01    \\
0.2     &       307.22  &       0.12    &       0.04    \\
0.3     &       298.08  &       0.27    &       0.09    \\
0.4     &       298.08  &       0.60    &       0.20    \\
0.5     &       301.86  &       1.07    &       0.35    \\
0.6     &       307.42  &       1.70    &       0.55    \\
0.8     &       321.15  &       3.13    &       0.98    \\
1.0     &       336.35  &       4.87    &       1.45    \\
1.2     &       352.61  &       7.80    &       2.21    \\
1.4     &       369.96  &       13.64   &       3.69    \\
1.6     &       387.97  &       23.40   &       6.03    \\
\multicolumn{4}{c}{He$_3$} \\
0.0     &       0.24    &       0.01    &       4.19    \\
0.2     &       0.40    &       0.01    &       3.57    \\
0.4     &       1.07    &       0.08    &       7.12    \\
0.6     &       2.68    &       0.21    &       7.14    \\
0.8     &       5.74    &       0.28    &       4.61    \\
1.0     &       10.66   &       0.25    &       2.31    \\
1.2     &       17.69   &       0.26    &       1.44    \\
1.4     &       26.87   &       0.61    &       2.20    \\
1.6     &       38.04   &       1.58    &       3.98    \\
1.8     &       50.89   &       3.34    &       6.16    \\
2.0     &       65.00   &       6.09    &       8.56    \\
\hline
\hline
\end{tabular*}
\end{footnotesize}
\end{table}
Note however, that care has to be taken when calculating BSSE corrected results as due to the artificial symmetry breaking in the magnetic field, 
state mixing may take place such that the comparison becomes meaningless. 
When comparing the binding energies obtained at the CCSD(T) level with HF results, we note that correlation increases the binding energies significantly by a roughly constant factor between 1.4 and 1.7. 

\begin{figure}
   \includegraphics{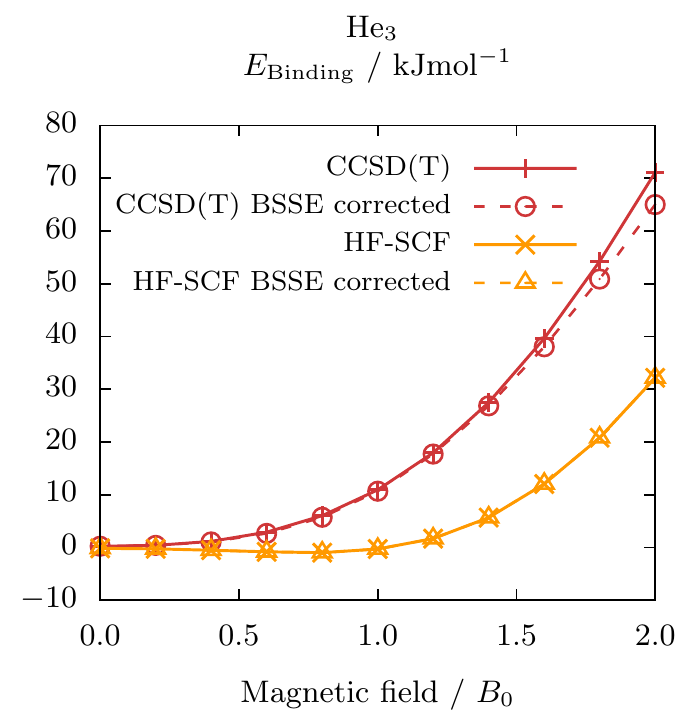}
   \caption{Binding energy for the singlet state of He$_3$ in perpendicular orientation as a function of the magnetic field. Calculated at the CCSD(T) (red) and HF-SCF level (orange) with and without BSSE correction (dotted lines and full lines, respectively) using the uncontracted aug-cc-pCVQZ basis set. }
   \label{He3_bonden}
\end{figure}

The picture looks different when turning to He$_3$ (see Figure \ref{He3_bonden}), which shows strong paramagnetic bonding in the magnetic field. 
While the system is unbound at the HF level until 1.0\,$B_0$ (note that it becomes bound earlier 
when geometries are optimized at the HF rather than CCSD(T) level \cite{ErikGradient}), calculations at the CCSD(T) level predict a weak van-der-Waals bonding already in the field-free case. 

We note that the binding energies for HF and CCSD(T) are not parallel. 
There is a much stronger increase in the binding energy in the correlated treatment such that the correlation contribution to the paramagnetic bonding is very large.
Consideration of electron correlation is therefore of great importance in the magnetic field.
This finding is in contrast to previous expectations about the importance of the correlation contribution to binding properties based on computational results for H$_2$.\cite{Schmelchercorr}
Concerning the BSSE (see Table \ref{table_BSSE}), we note that the error is, as expected for a van-der-Waals bound system, already quite large for the zero-field case---that is, 4.2$\%$. 
As the magnetic field increases, the BSSE increases as well, but with the relative error behaving in a more complicated manner due to the interplay of van-der-Waals and increasing paramagnetic bonding.
\section{conclusion}
We have presented a first implementation of the CCSD and CCSD(T) schemes for the quantum-chemical treatment of molecules in magnetic fields beyond the perturbative limit and beyond magnetic fields observable on Earth. 
Our findings show that correlation energies for atoms and particularly for molecules behave in a complex manner. 
While, for atoms, the change in correlation energy is mostly determined by the confinement effect---that is, 
according to whether the state contracts or expands in the correlated treatment relative to the HF level---the situation for molecules is more involved as equilibrium geometries change as a function of the field, making it necessary to investigate potential energy surfaces.
Additionally, for molecules there are fewer symmetry constraints such that the response in the magnetic field becomes more complicated. 
Basis-set requirements are also much more pronounced for the treatment in magnetic field, especially if states of higher angular momentum need to be described. 
As such states are very likely to become ground states in the presence of the field, their proper description is much more important than for field-free calculations.
\red{
Our study of basis-set convergence in finite magnetic fields and the comparison to literature data obtained with anisotropic Gaussians reveals that with uncontracted standard basis sets of quadruple-zeta quality, results can typically be trusted up until 1--1.5\,$B_0$.
}
While basis functions with anisotropic Gaussians can at least deal with the issue of deformation of orbitals in the magnetic field, describing states of higher angular momentum will still remain an issue. 
For binding energies, the effect of electron correlation can be very different as illustrated by LiH and He$_3$. 
While for the former, electron correlation introduces a more or less constant shift to higher binding energies, for the latter the correlated treatment predicts a much steeper increase in paramagnetic bonding.

The work presented enables for the first time a thorough investigation of
electron-correlation effects in atoms and molecules in the presence of
strong  magnetic fields, emphasizing the importance of including electron correlation in
corresponding quantum-chemical treatments.
Coupled-cluster theory provides here an ideal vehicle, but its application
in its standard formulation is restricted to systems that are well
described by a single Slater determinant. 
Despite this restriction and the increasing importance of open-shell systems with multireference character in the presence of magnetic fields, the coupled-cluster implementation presented in this work already allows the investigation of many interesting systems and their electronic
states in finite magnetic fields and, for example, provides important benchmark data
for the development of appropriate density functionals to deal with systems in magnetic
fields.\cite{Andy_CDFT} 
The formulation and implementation of an equation-of-motion \cite{EOM} ansatz would represent a useful extension of the present work, enabling the treatment of systems with multi-reference character as well as, for example, the additional computation of excitation spectra. 
A further possible extension of the present work could deal with molecular properties in the presence of magnetic fields and would involve the implementation of analytic energy derivatives \cite{molprop} as well as corresponding response-theory approaches.\cite{response} 
\\
\section{Acknowledgments}
Financial support in Oslo by the Norwegian Research Council through the CoE Centre for Theoretical and Computational Chemistry 
(Grant Nos. 179568/V30 and 197446/V30),   
through the European Research Council under the European Union Seventh Framework Program through the Advanced Grant ABACUS, ERC Grant Agreement No. 267683, 
and in 
Mainz by the Deutsche Forschungsgemeinschaft (DFG GA-370/5-1) is gratefully acknowledged.

In addition, J.G. also thanks the Centre of Theoretical and
Computational Chemistry (CTCC) at the University of Oslo
for the hospitality and financial support during a sabbatical. 
\newpage

%

\end{document}